\documentclass{article}

\usepackage[round]{natbib}

\usepackage[utf8]{inputenc} 
\usepackage[T1]{fontenc}    
\usepackage{hyperref}       
\usepackage{url}            
\usepackage{booktabs}       
\usepackage{amsfonts}       
\usepackage{nicefrac}       
\usepackage{microtype}      
\usepackage{xcolor}         

\usepackage{amsmath}
\usepackage{cblocker}
\usepackage[pdftex]{graphicx}
\usepackage{wrapfig}
\usepackage{caption}
\usepackage{subcaption}

\providecommand{\eqnref}[1]   {(\protect\ref{#1})}

\newtheorem{theorem}{Theorem}

\newcommand{\laura}[1] {{\color{black}#1}}

\newcommand{\new}[1] {{\color{black}#1}}

\title{Dynamic Subspace Estimation with Grassmannian Geodesics}

\newcommand{\email}[1]{\href{mailto:#1}{#1}}

\author{
Cameron J.~Blocker\thanks{Department of Electrical Engineering and Computer Science, University of Michigan, Ann Arbor. (\email{cblocker@umich.edu}).}
\and Haroon Raja\thanks{Eli Lilly, Indianapolis, IN (\email{raja\_haroon@lilly.com}).}
\and Jeffrey A.~Fessler\thanks{Department of Electrical Engineering and Computer Science, University of Michigan, Ann Arbor. (\email{fessler@umich.edu}).}
\and Laura Balzano\thanks{Department of Electrical Engineering and Computer Science, University of Michigan, Ann Arbor, USA. (\email{girasole@umich.edu}.)}
}

\begin{document}

\maketitle

%

%

%






\vspace{5mm}
\begin{abstract}
    Dynamic subspace estimation, or subspace tracking, 
    is a fundamental problem in
    statistical signal processing and machine learning. 
    This paper considers a geodesic model 
    for time-varying subspaces. The natural objective function for this model is non-convex.
    We propose a novel algorithm for minimizing this objective and estimating the parameters of the model 
    from data with Grassmannian-constrained optimization. We show that with this algorithm, the objective is monotonically non-increasing. 
    We demonstrate the performance of this model and our algorithm on synthetic data, video data, and
    dynamic fMRI data.
\end{abstract}

\section{Introduction}

Modeling data using linear subspaces 
is a powerful analytical tool 
that enables practitioners to 
more efficiently and reliably 
solve high-level tasks 
like inference and decision making, 
classification, and anomaly detection, 
among others. 
In some applications of interest, 
the data generation process is time-varying or dynamic in nature, 
which motivates the use of a dynamic linear subspace for data modeling. 
Some example applications 
where dynamic subspace models are prevalent 
include array signal processing~\citep{
    yang1995projection,
    fuhrmann1997geometric,
    srivastava2004bayesian,
    lake1998maximum},
communication systems~\citep{haghighatshoar2018low}, 
video processing~\citep{vaswani2018robust},
and dynamic magnetic resonance imaging (MRI)~\citep{otazo2015low}.
The goal in these applications 
is to learn a time-varying subspace from the observed data. 

Most previous theoretical work 
for modeling a dynamic subspace 
relies on very strong assumptions of the dynamics 
-- either assuming very simple dynamics 
like sudden changes with an otherwise static subspace, 
or assuming a specific known dynamical model. 
A much broader empirical literature for \emph{subspace tracking} 
considers a wide range of algorithms 
with different strengths and weaknesses 
with regards to signal-to-noise ratios, 
speed of dynamics, and computational complexity. 
For the vast majority of these algorithms,  accuracy guarantees in the presence of dynamics 
are still an open question. 

This paper starts with a flexible and natural dynamic subspace model: 
the piecewise geodesic model. A piecewise geodesic can approximate any curve on the Grassmannian,
i.e., any continuously varying subspace.
This model generalizes both 
the previously studied time-varying subspace models
and piecewise linear approximations 
that are pervasive in the theory and practice 
of statistical signal processing. 
This model has only been very briefly discussed 
in existing literature, 
probably in part due to the difficulty of parameter estimation
and algorithmic guarantees in this setting. 
In this paper, we start by learning a single geodesic.
The central contribution of this paper, therefore, is an algorithm for learning the parameters 
of this model in a batch setting that is guaranteed to descend an appropriate cost function
(corresponding to a log-likelihood for Gaussian noise)
at every step. 
We also demonstrate the performance of the proposed algorithm 
empirically on both synthetic and real datasets.


\subsection{Problem Formulation and Geodesic Model}  \label{sec:model}

We start with the following broad generative model 
for data arising from a time-varying subspace. 
At each time point \(i\) 
we observe \(\ell\) vectors from a time-varying subspace.
Let \(\X_i\in \reals^{d \times \ell}\) for \(i=1,2,\dots,T\) 
be data generated from a low-rank model with noise:
\begin{equation}\label{eqn:gen}
    \X_i = \U_i \G_i + \N_i
\end{equation}
where \(\U_i \in \reals^{d \times k}\) is a matrix with orthonormal columns representing a point on the Grassmannian $\grass(k,d)$, the space of all rank-$k$ subspaces in $\reals^{d}$;
\(\G_i \in \reals^{k \times \ell}\) holds weight or loading vectors;
and \(\N_i \in \reals^{d \times \ell}\) is 
an independent additive noise matrix. 
We observe \(\X_i\) and our objective is 
to estimate \new{a sequence of subspaces \(\{\U_i\}_{i=1}^T\) that generates the observed samples.} \(\U_i\) for \(i=1,\dots,T\). Note that while we use ``time-varying'' to describe this generative model, it can vary over any index, and the algorithms we consider are batch in the sense that they use all the data \(\X_i\), \(i=1,\dots,T\) for estimating \(\U_i\), \(i=1,\dots,T\).

\begin{figure}
\centering
	\includegraphics[width=0.6\columnwidth]{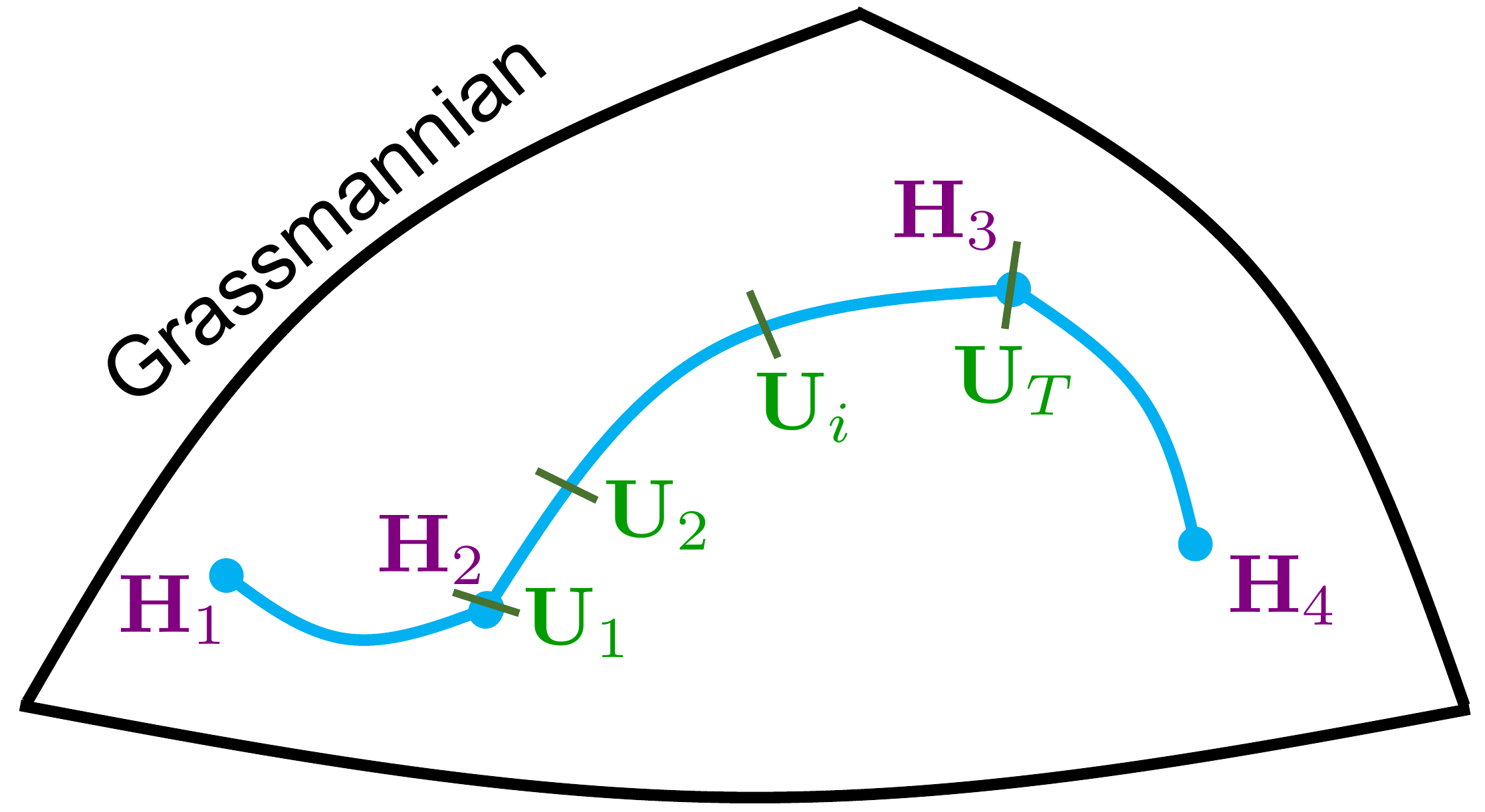}
		\caption{
		\raggedright
		Illustration of the piecewise geodesic model.
		$\mathbf{H_1}, \dots, \mathbf{H_4}$ are points on the Grassmannian.
		When estimating a single geodesic, e.g.,
		the one from $\mathbf{H_2}$ to $\mathbf{H_3}$,
		then \H is an orthonormal basis for $\mathbf{H_2}$
		and \Y is an orthonormal basis for $(\I-\H\H^\top) \mathbf{H_3}$.
		}
\label{fig:piecewisegeod}
\end{figure}

If \(\U_i = \U^|\) is static for all \(i = 1, 2, \dots, T\), 
and if \G_i has zero-mean columns, 
then we could concatenate all \(\X_i\) together and apply the SVD, which is well-known to recover a good approximation of \U^| 
as long as the number of samples \(\ell T\) 
is large enough to overcome the noise. 
However, if \U_i is varying for every $i$, 
one can immediately see that if \(\ell < k\), 
estimating \U_i is impossible without further assumptions. 
Even when \(\ell \geq k\), 
in many applications
it is natural
to impose a relationship 
between the \U_i subspace matrices over time, 
to guarantee regularity properties 
or known application constraints. 
Various constraints have been studied in the literature, 
such as a slowly rotating subspace, 
a subspace that is mostly static except for intermittent sudden changes, 
or a subspace that changes one dimension at a time~\citep{narayanamurthy2018provable}.
Those models are all subsumed 
by the \textbf{piecewise geodesic} model for dynamic subspaces,
illustrated in Figure~\ref{fig:piecewisegeod}. 
In this work,
we focus on efficiently learning a single geodesic.


\paragraph{Model for a Single Geodesic}

Let $2k \leq d$. We model each {\U_i} 
as an orthonormal basis whose span has
been sampled from a single continuous Grassmannian geodesic 
\(\U(t):~[0,1]\rightarrow \stief^{d\times k}\)
parameterized as follows:
\begin{align}
    \U_i &= \U(t_i)
    = \H \cos{\bTheta t_i} + \Y \sin{\bTheta t_i}
    \label{eqn:model}
\end{align} 
where 
\stief^{d\times k} is the set of \(d\times k\) matrices 
with orthonormal columns (the Stiefel manifold), 
\(\H \in \stief^{d\times k}\) is an orthonormal basis
for a point on the Grassmannian, 
\Y is a matrix with orthonormal columns
whose span is in the tangent space of the Grassmannian 
at \Span{\H}, 
\ie, \(\Y \in \set{\Y | \H^\top\Y=\mathbf{0}, \Y \in \stief^{d\times k}}\),
and \( \bTheta \in \reals^{k \times k} \) is 
a diagonal matrix 
where $\theta_j$ is the $j$th principal angle between the two endpoints of the geodesic, and sine/cosine are the matrix versions.
These constraints
ensure each \U_i
has orthonormal columns.
The scalars
\(t_i \in [0,1]\) represent the location of each $\U_i$ along the geodesic,
\eg,
if the geodesic is sampled over time, these are time-points scaled (or normalized) to the interval. 
For more information, see \cite[Section~3.8]{absil2004riemannian} and \cite{edelman:98:tgo}. 

Because we are only interested in the span of \U_i,
this parameterization of a Grassmannian geodesic is not unique.
Permuting the columns of \H, \Y and the diagonal elements of \bTheta
would result in a \U_i with the same span.
Additionally, there is a sign ambiguity between columns of \Y and diagonal elements of \bTheta.
In practice, 
our loss is invariant to these ambiguities 
and so they are not a problem.
Any specific parameterization
can easily be transformed into another.


\H, \Y and \bTheta are all learnable parameters of \(\U(t)\).
Conceptually, we can think of \H as a starting point on the Grassmannian, 
\Y as a normalized direction we want to walk, 
and the diagonal elements of
\(\bTheta t_i\)
as the distances in each dimension we should walk
from \H
on the surface of the manifold to get to \U_i.

\paragraph{Single Geodesic vs. Piecewise Geodesic} In this work, we focus on learning a single geodesic 
from data with given time points $t_i$.
This focus essentially makes two key simplifying assumptions: 
(1) the locations of the knots, or change-points, in a piecewise approximation are given,
and 
(2) between two knots in the piecewise approximation, either the time-points $t_i$ are given,
or observed matrices $\X_i$ are equidistant along a geodesic curve. 
With these assumptions, our high-level approach is to take each set of data matrices between change-points
and learn a single geodesic.
Section \ref{apx:addlossi} discusses how to learn multiple geodesics with knots known.
We plan to relax both of these assumptions in future work. 

\subsection{Related work}

Classical literature on subspace tracking 
uses online approaches to estimate 
the time-varying subspaces~\citep{
     yang1995projection,
     chi2013petrels,
     allen2017follow,
     balzano2018streaming,
     haghighatshoar2018low,
     narayanamurthy2018provable,
     vaswani2018robust,
     comon1990tracking}.
Early theoretical results were limited 
to asymptotic convergence guarantees 
with static underlying subspaces. 
Among the more recent works, 
the PETRELS algorithm~\citep{chi2013petrels} 
portrays a recursive least squares approach 
and provides convergence theory that assumes 
that the subspace changes at a particular instant 
and then stays constant for sufficient time 
so that the change can be tracked 
(also called the piecewise constant model). 
\cite{narayanamurthy2018provable} 
relax the assumption of constant subspace 
to a very slowly varying subspace between the change points.
For a review of these methods, see%
~\cite{vaswani2018robust}.

Dynamic subspace estimation 
has also been studied for the more general Grassmannian geodesic model that is the focus of this paper~\citep{
    lake1998maximum,
    fuhrmann1997geometric,
    srivastava2004bayesian,
    hong2016parametric}.  
Unlike the subspace tracking problem,
these contributions focus on batch data settings with access to the whole dataset for estimation,
which is the approach we take herein. 
For example, \cite{lake1998maximum} 
formulate the subspace tracking for any given epoch 
in the generative geodesic model
in the form of a likelihood function to maximize. 
The likelihood is nonconcave, 
but the authors provide an annealing approach to solve it, 
which will have very high computational burden 
for many modern high-dimensional and large-data applications.
\cite{fuhrmann1997geometric} 
and \cite{srivastava2004bayesian}
have also studied the generative geodesic model. 
The solution provided by 
\cite{srivastava2004bayesian} 
is not applicable to large-scale settings 
since it relies on a sampling based strategy 
like Markov chain Monte-Carlo (MCMC) 
and hence is computationally intensive. 
On the other hand, 
the solution provided by \cite{fuhrmann1997geometric}
is computationally inexpensive, 
but it only handles
one-dimensional 
subspaces (\(k=1\)).
In summary,
major weaknesses 
of the state-of-the-art methods 
include high computational costs,
lack of theoretical guarantees, and/or need to tune hyperparameters. 
This paper approaches the problem using modern nonconvex optimization tools,
alleviating these issues 
by devising an algorithm that is parameter free
other than the subspace dimension,
descends the loss function monotonically,
and uses thin
SVDs to solve the problem. 

\section{Method}

Given data and associated time points \set{\X_i, t_i}_{i=1}^T, 
we fit the proposed geodesic model for \(\U(t)\) by minimizing the following loss function
\begin{align}
    \loss{\U}
    &= \loss(\H, \Y, \bTheta) \\
    &= \min{\set{\G_i}_{i=1}^T} \sum_{i=1}^T \normF{\X_i - \U(t_i)\G_i}^2 \\
    &= -\sum_{i=1}^T \normF{\X_i^\top\U(t_i)}^2 + c \label{eqn:loss2}
,\end{align}
where in the last equality we have substituted 
the optimal \( \G_i = (\U(t_i))^\top\X_i\) and simplified
and $c$ is a constant \citep{golub:03:snl}.
At a high level,
we minimize our loss function
with respect to \H, \Y, and \bTheta
via block coordinate descent.
Our first block updates \((\H,\Y)\)
jointly via an SVD of a \(d\times 2k\) matrix.
Next, we update \bTheta via a first-order iterative minimization.
Specifically, we design an efficient majorize-minimize (MM) iteration
for updating \bTheta.

\subsection{\texorpdfstring{(\H, \Y)}{(H,Y)} Update}

\def\iteridx {n}

Let \(\Q \defequ [\H\ \Y]\) and \(\Z_i \defequ [\cos{\bTheta t_i} ; \, \sin{\bTheta t_i} ]\).
Then we rewrite our model \eqnref{eqn:model}
as \(\U_i = \Q\Z_i\).
By constraining
\(\Q \in \stief^{d \times 2k}\),
we also satisfy the constraints of \H and \Y individually.

Following
\cite{breloy:21:mmo},
we form a linear majorizer for our loss \eqnref{eqn:loss2}
at \(\Q\iterk\)
and minimize it with a Stiefel constraint
simply by projecting its negative gradient onto the Stiefel manifold.
The update is then given by
\begin{align}
    \Q\iterk[+1] &= \argmin{\Q\in\stief^{d\times 2k}}
    \normF{\Q - \sumno_{i=1}^T \X_i\X_i^\top\Q\iterk\Z_i\Z_i^\top}^2
    = \W\V^\top
    \label{eqn:svdx}
,\end{align}
where $\W\bSigma\V^\top$ is the SVD of
$\sum_{i=1}^T \X_i\X_i^\top\Q\iterk\Z_i\Z_i^\top$.
Further, using the simplification from \eqref{eqn:loss2} we have
\(\hat{\G}_i^\top = \X_i^\top\U_i\iterk = \X_i^\top\Q\iterk\Z_i\).
Using this equality we can express the summation term in \eqnref{eqn:svdx}
in terms $\X_i$, $\hat\G$, and $\bTheta$ as follows
\begin{align*}
    \sum_{i=1}^T \X_i\X_i^\top\Q\iterk\Z_i\Z_i^\top =
    \sum_{i=1}^T \begin{bmatrix} \X_i\hat{\G_i}^\top\cos{\bTheta t_i} & \X_i\hat{\G_i}^\top\sin{\bTheta t_i}\end{bmatrix}
,\end{align*}
where 
we let 
\(\hat{\G}_i^\top = \X_i^\top\U_i\iterk = \X_i^\top\Q\iterk\Z_i\).
Although here we eliminated \set{\G_i}_{i=1}^T from our loss \eqnref{eqn:loss2},
if we had not, this same update could be derived as 
a block coordinate update on \Q and \set{\G_i}_{i=1}^T.
See Appendix~\ref{apx:HY_mm} for more details.

\subsection{\texorpdfstring{\bTheta}{theta} Update}

The loss \eqnref{eqn:loss2}
for fixed \H and \Y
is a smooth function of \bTheta 
and can be effectively optimized 
via an iterative quadratic majorize-minimize scheme.
Here we provide an overview of the method.
Appendix~\ref{apx:theta_mm}
provides a more complete derivation of the majorizer.
First, we simplify the loss \eqnref{eqn:loss2}
\begin{align*}
    \hat\bTheta &= \argmin{\bTheta}
                -\sum_{i=1}^T \normF{\X_i^\top\U_i}^2 \\
                &= \argmin{\bTheta} 
                -\sum_{i=1}^T \normF{\X_i^\top\left(\H\cos{\bTheta t_i}+
                                            \Y\sin{\bTheta t_i}\right)}^2 \\
            &= \argmin{\bTheta} 
                -\sum_{i=1}^T \sum_{j=1}^k  r_{i,j}\cos(2\theta_j t_i - \phi_{i,j}) + b_{i,j},
\end{align*}
where defining \(\arctan\!2(y,x)\) as the angle 
of the point \((x,y)\) 
in the 2D plane 
counter-clockwise from the positive \(x\)-axis, the associated constants
\(r_{i,j}, \phi_{i,j}, b_{i,j}\)
are defined as
\begin{align}
    \phi_{i,j} &= \arctan\!2\!\left(\beta_{i,j},\frac{\alpha_{i,j}-\gamma_{i,j}}{2}\right) \\
    r_{i,j} &= \sqrt{ {\left( \frac{\alpha_{i,j}-\gamma_{i,j}}{2} \right)}^2 
                  + \beta_{i,j}^2 } \\
    b_{i,j} &= \frac{\alpha_{i,j} + \gamma_{i,j}}{2}\\
    \alpha_{i,j} &= {[\H^\top\X_i\X_i^\top\H]}_{j,j} \label{eq:calcalpha}\\
    \beta_{i,j} &= \real{{[\Y^\top\X_i\X_i^\top\H]}_{j,j}} \label{eq:calcbeta}\\
    \gamma_{i,j} &= {[\Y^\top\X_i\X_i^\top\Y]}_{j,j}\label{eq:calcgamma}
.\end{align}
This loss is separable 
for each diagonal element of \bTheta,
so we find each $\hat\theta_j$
via a (1D) minimization.
Let \(\f_{i,j}(\theta_j) \defequ -r_{i,j}\cos(2\theta_j t_i - \phi_{i,j}) + b_{i,j}\).
Then
(cf.~\cite{funai:08:rfm})
the following \q_{i,j} defines 
a quadratic majorizer for \f_{i,j} at \theta_j
\begin{align}
    \q_{i,j}(\theta_j; \theta_j^|) &= \f_{i,j}(\theta_j^|) 
                                + \f_{i,j}^.(\theta_j^|)(\theta_j-\theta_j^|)
                                + \frac12\w_{\f_{i,j}}(\theta_j^|){(\theta_j-\theta_j^|)}^2
\nonumber\\&
    \geq \f_{i,j}(\theta_j)
\end{align}
where the derivative \f_{i,j}^. and curvature function \(\w_{\f_{i,j}}\) are given by
\begin{align*}
    \f_{i,j}^.(\theta_j) &= 2r_{i,j}t_i \sin{2\theta_j t_i - \phi_{i,j}} \\
    \w_{\f_{i,j}}(\theta_j) &= \frac{
               \f_{i,j}^.(\theta_j)}{ 
               \mathrm{mod}\!\left(
                    (\theta_j - \frac{\phi_{i,j}}{2t_i}) + \frac{\pi}{2t_i}, 
                    \frac{2\pi}{2t_i}
               \right) - \frac{\pi}{2t_i} }
.\end{align*}
Appendix~\ref{apx:theta_mm:maj}
gives a detailed construction of \(\w_{\f_{i,j}}\).


Our majorize-minimize iterations
for each diagonal element of \bTheta
are then given by
\begin{align}{
    \theta_j\iterk[+1] &= \argmin{\theta_j} \sum_{i=1}^T 
                             \q_{i,j}(\theta_j; \theta_j\iterk)
\\&
    = \theta_j\iterk - 
        \frac{\sum_{i=1}^T \f_{i,j}^.(\theta_j\iterk) }
            {\sum_{i=1}^T \w_{\f_{i,j}}(\theta_j\iterk)}
.}\end{align}

\vspace{-1mm}
Conceptually, each MM update can be interpreted
as a gradient descent step with a variable step size \(s\iterk = {1/(\sum_{i=1}^T \w_{\f_{i,j}}(\theta_j\iterk))}\)
that is guaranteed to not increase the loss
even without any line search.
Indeed, as both updates just outlined are known for their monotonicity properties, we have the following monotonicity result for our overall algorithm. 
\laura{
Since the algorithm
monotonically decreases the loss, which is bounded, this implies that the loss always converges.
Empirically we see
convergence to the planted subspace end points in the vast majority of experiments, but not in all.
A more thorough investigation of the loss landscape and algorithmic convergence properties
is of great interest for future work. }

\begin{theorem}\label{thm:mono}
\raggedright Algorithm~\ref{alg:gse} produces iterates ${\U\iterk(t) = \H\iterk \cos(\bTheta\iterk t)} + \Y\iterk \sin(\bTheta\iterk t)$ 
that are monotonically non-increasing in loss~\eqnref{eqn:loss2},
i.e.,
\(
    \loss{\U\iterk[+1]} \leq \loss{\U\iterk}
\).
\end{theorem}

\emph{Proof:} It suffices to show that each block coordinate update
does not increase the loss.
Both the \([\H\ \Y]\) block and \bTheta block updates are instances
of MM methods that guarantee this property.
See, for example, \cite{sun:17:mma} Section II.C
for a general treatment
and \cite{breloy:21:mmo} Section III.B
for MM convergence with the nonconvex Stiefel constraint.

\laura{
\paragraph{Complexity}
The time complexity for each iteration is $O(T dk\ell)$, assuming $T\ell \geq k$. This is the complexity for both the $\Q$
and $\bTheta$ update. 

For the \Q update we need to form the matrix $\sum_{i=1}^T \X_i\X_i^\top\Q\iterk\Z_i\Z_i^\top$ and take its SVD (see \eqref{eqn:svdx} and the surrounding text). This requires $O(T d k \ell + dk^2)$ operations, where $O(T d k \ell)$ forms the sum of the product of matrices and $O(dk^2)$ computes the SVD of this $d \times 2k$ matrix.}



\laura{For the $\bTheta$ update, the key computational steps form the matrix products $\H^\top \X_i$ and $\Y^\top \X_i$ that are used in the calculation of $\alpha, \beta, \gamma$ from \eqref{eq:calcalpha}, \eqref{eq:calcbeta}, \eqref{eq:calcgamma}, which require $O(dk\ell)$ operations for each $i=1,\dots, T$. The remaining computations involve scalar operations on $k$ or $Tk$ variables, and thus the overall computational complexity is $O(Tdk\ell)$.}

\section{Experiments}

To show the effectiveness of the proposed method,
we present results on both synthetic and real data.
On the synthetic data,
we show the effect of the different data parameters,
such as \(d, k, \ell, T\) and the additive noise standard deviation \sigma.
With the intuition we build on the synthetic data,
we present results on real measured data
and show how we determine the underlying rank.
All experiments were performed
on a 2021 Macbook Pro laptop computer
and implemented in Python.%
\footnote{
    Code and data will be open sourced with MIT license.
}

\subsection{Synthetic Data}

\begin{figure}
  \centering
   \includegraphics[width=.8\linewidth]{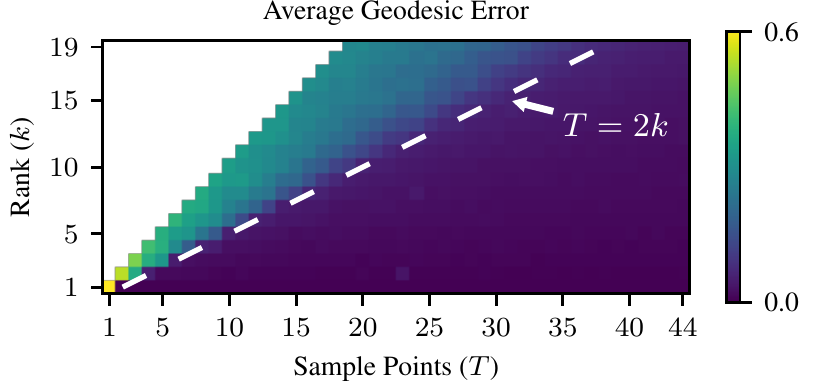}
  \caption{The average geodesic error over 15 trials for
  varying rank \(k\) and number of sample points \(T\).
  One vector was sampled at each of \(T\) points (\(\ell = 1\)).
  The ambient dimension was \(d = 40\),
  and we added zero-mean white Gaussian
  noise with standard deviation \(\sigma = 10^{\mhyphen5}\).
  We see a phase transition at \(T=2k\); with at least this many samples, 
  we recover the true subspace with low error.
  }\label{fig:phase}
\end{figure}

\begin{figure}
  \centering
   \includegraphics[width=.6\linewidth]{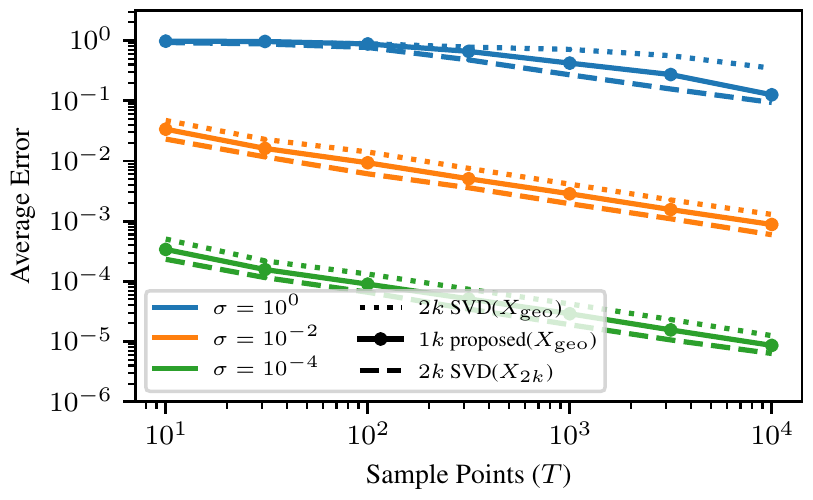}
  \caption{The recovered geodesic error (solid lines) as a function of sample size and additive noise standard deviation, 
  averaged over 10 problems.
  \(\X_{\mathrm{geo}}\) represents the batched geodesic data
  that has data distributed like a geodesic in a rank-$2k$ subspace.
  \(\X_{2k}\) denotes data that is distributed isotropically 
  in a rank-$2k$ subspace.
  For data generated from a geodesic,
  the proposed method recovers the geodesic error with a lower error than an
  SVD can estimate its span.
  }\label{fig:error_vs_samples}
\end{figure}

In the case of synthetic data,
we are able to compare our estimated geodesic \(\hat{\U}(t)\)
against the true geodesic
from which the data was generated.
Our error metric is the square root of the
average squared subspace error between corresponding points along the geodesic
\begin{align}
    \label{eqn:ssp_error}
    \text{Subspace Error} &= \frac1{\sqrt{2k}} \normFr{\hat{\U}\hat{\U}^\top - \U\U^\top} \\
    \label{eqn:geo_error}
    \text{Geodesic Error} &= 
    \sqrt{\int_0^1{\frac1{2k} \normFr{\hat{\U}(t)\hat{\U}(t)^\top - \U(t)\U(t)^\top}^2 \der t}}
\ .\end{align}
In practice, we approximate the integral by sampling the geodesic at a large number of time points.
The subspace error \eqnref{eqn:ssp_error} takes a minimum value of 0 when
\(\mathrm{span}(\hat{\U})=\Span{\U}\) 
and a maximum value of 1 when \(\mathrm{span}(\hat{\U})\perp\Span{\U}\).
As such, the geodesic error \eqnref{eqn:geo_error} is similarly bounded.
For all of our synthetic experiments,
we generated data from our planted model
\eqref{eqn:gen}.
\bTheta was constrained
to generate distance-minimizing geodesics,
\ie, \(\opnorm{\bTheta}_2 < \nicefrac{\pi}{2}\).
We drew
\G_i from a standard normal distribution.
The noise \N_i is additive white Gaussian noise (AWGN)
with standard deviation \sigma.
Unless otherwise noted,
we initialized the proposed method
with a random geodesic
for all experiments.
For the experiments
in Figure~\ref{fig:phase} and
Figure~\ref{fig:error_vs_samples},
we formed a coarse estimate
of the starting and ending
rank-$k$ subspaces, and
computed a geodesic between
these subspaces with~\cite[(19)]{absil2004riemannian}
for initialization.

Figure~\ref{fig:phase} shows the average geodesic error when $\ell=1$, i.e.,
we receive one vector per $k$-dimensional subspace,
as a function of both the true underlying rank (\(k\))
and the number of sample points (\(T\)).
This plot shows that a phase transition occurs at \(T=2k\),
where if we have at least this many samples, 
the proposed method can recover the true geodesic with low error.
At least $2k$ samples are necessary to compute the rank-$2k$ SVD,
and this figure shows that $2k$ samples are also sufficient for computing the geodesic endpoints.

In Figure~\ref{fig:error_vs_samples},
we further investigate the effect of the number of samples on the average geodesic error.
Because a rank-$k$ geodesic spans a space as large as $2k$,
we have also shown for reference the subspace error of recovering a rank-$2k$
subspace with an SVD under two different distributions of loading vectors.
The dashed lines show the subspace recovery error for data generated
isotropically in a rank-$2k$ subspace with additive white Gaussian noise.
The dotted lines show the subspace recovery error for data distributed on
a geodesic in the rank-$2k$ subspace.
Both SVD-based methods are only recovering a single rank-$2k$ subspace
and not recovering a geodesic.
Empirically, we can see that the sample complexity of the proposed geodesic model and method tracks
well with rank-$2k$ SVD and outperforms SVD on geodesic data.

\begin{figure}
  \centering
   \includegraphics[width=.6\linewidth]{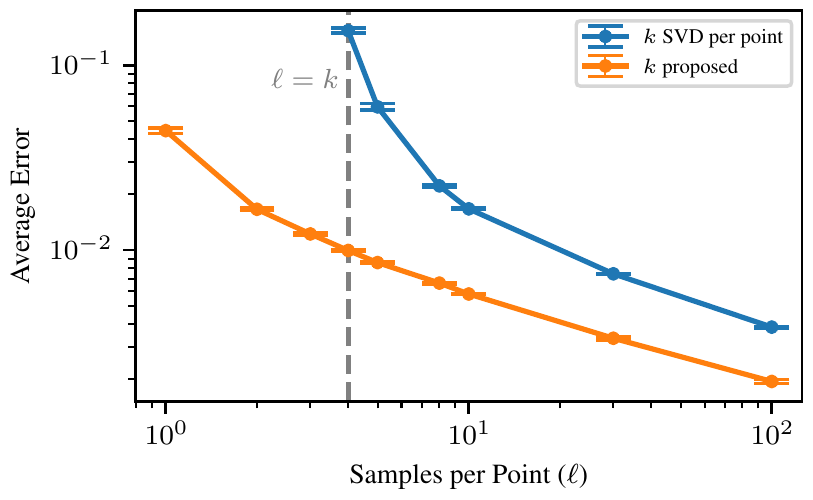}
  \caption{Average geodesic error over 100 trials, \laura{with standard error bars,} for varying
  number of samples (\(\ell\)) collected at each time point for
  a fixed number of time points ($T=11$) on a planted rank-4 geodesic
  with AWGN \(\sigma = 10^{\mhyphen2}\).
  When \(\ell \geq k\) we can estimate the subspace with
  the SVD on just those \(\ell\) samples.
  The geodesic model can estimate the subspaces even when
  \(\ell < k\) and leverages all of the data to produce lower error.
  }\label{fig:error_vs_ell}
\end{figure}


When the number of samples per time point (\(\ell\))
is less than \(k\)
the proposed method is still able to recover the true
geodesic given that \(T\) is large enough.
When \(\ell \geq k\),
one could estimate the subspace at each time point
by applying a rank-$k$ SVD at each time point.
But, as shown in
Figure~\ref{fig:error_vs_ell},
even when \(\ell \geq k\),
the proposed method recovers the subspaces
at each time point with lower error
for data generated from a geodesic.

\begin{figure*}
  \centering
   \includegraphics[width=\linewidth]{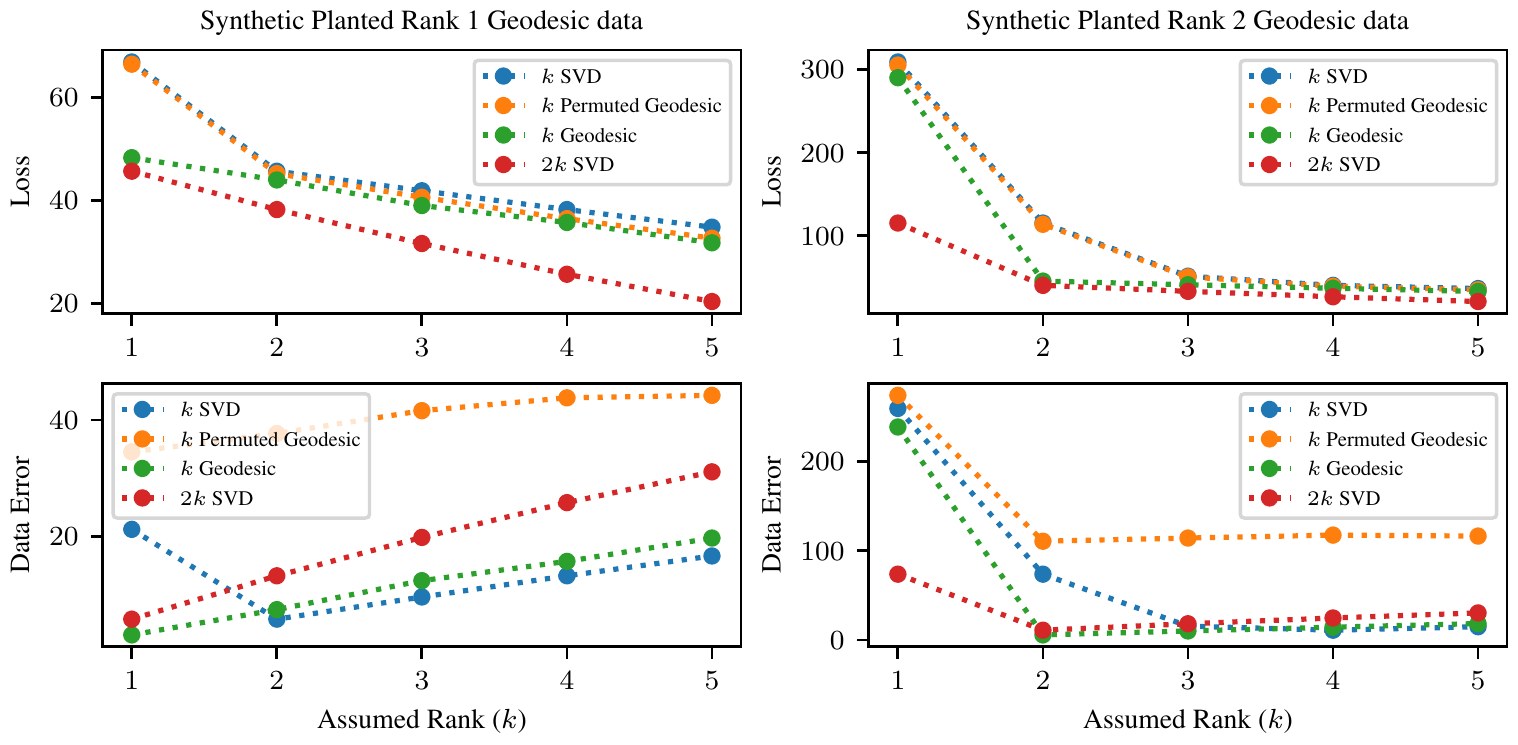}
  \caption{Loss on synthetic data for a rank-1 geodesic (left)
  and a rank-2 geodesic (right).
  The loss of the proposed method
  is lower-bounded and upper-bounded by rank-$2k$ and rank-$k$ SVD,
  respectively. When the assumed rank is equal to the true rank,
  than the loss of the proposed method is much closer to that
  of rank-$2k$ SVD, while permuting the data significantly increases the loss. 
  From this, it is easy to deduce the true rank.
  The second row shows the ``Data Error,''
  which is the norm of the residual between the projected noisy data and the noiseless data.
  We can see that rank-$2k$ SVD was overfitting noise to obtain a lower training loss,
  but this increases its error.
  The proposed method has a lower error than rank-$2k$ SVD 
  as long as the assumed rank is greater than or equal to the true rank
  on geodesic data.
  }\label{fig:syn_loss}
\end{figure*}

\begin{figure}
  \centering
   \includegraphics[width=0.6\linewidth]{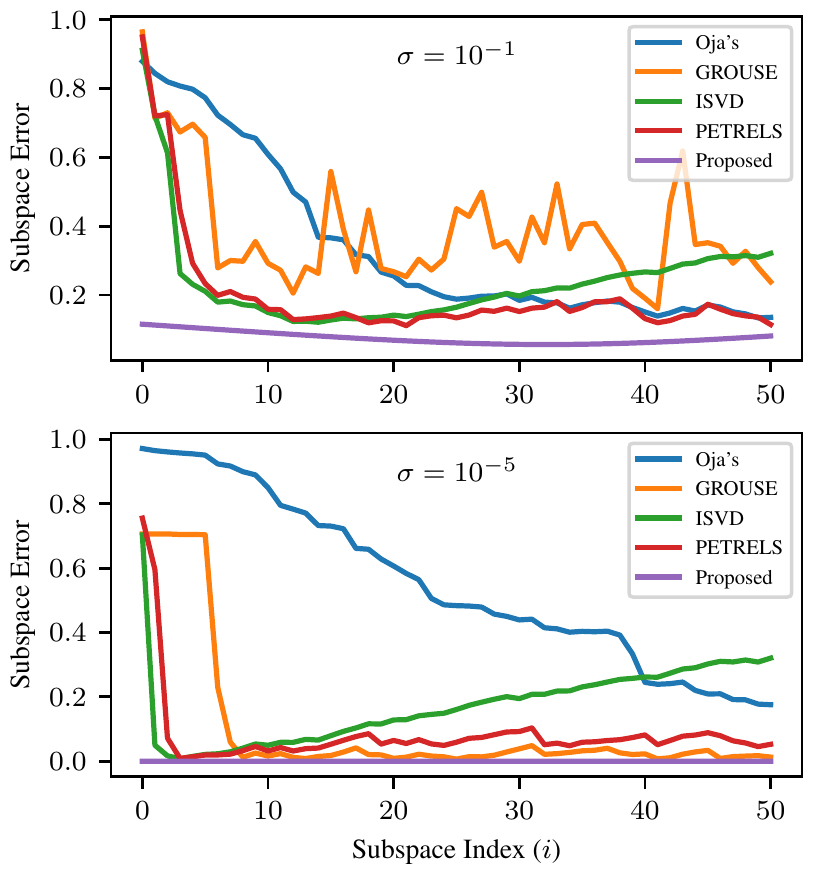}
  \caption{\new{Subspace error at each time point
  for the proposed method compared to 
  several online subspace tracking algorithms
  at two different noise levels \sigma.
  We sampled a single vector 
  from 51 time points
  of a rank-$2$ geodesic
  for both noise cases.
  Note that the online algorithms estimate a single
  time point at a time,
  while the proposed method considers all time points.}
  }\label{fig:streaming}
\end{figure}

Like a low-rank SVD approximation, our method requires choosing
the rank $k$ before fitting.
Figure~\ref{fig:syn_loss} shows the loss as a
function of assumed rank for data generated
from a rank-1 geodesic (left) and rank-2 geodesic (right).
Because a rank-$k$ geodesic spans a rank-$2k$ subspace,
rank-$k$ and rank-$2k$ SVD results are shown for comparison.
Rank-$2k$ SVD will always have a lower loss by definition.
Similarly, the proposed model will always have a lower loss
than the rank-$k$ SVD,
since a rank-$k$ subspace is a special case of a rank-$k$ geodesic.
Thus, we can lower-bound and upper-bound the loss
of the proposed model on any data by
a rank-$2k$ and rank-$k$ SVD respectively.
Additionally, if the data has geodesic structure,
then it is ordered.
For comparison we also show the loss
of the proposed method on data that was generated from a geodesic and then permuted.
We see that when the assumed rank is equal to the true rank
of the underlying geodesic, the proposed method produces a loss much closer to
that of a rank-$2k$ SVD
while the proposed method applied to permuted data
produces a loss much closer to that of a rank-$k$ SVD.
For permuted unordered data,
the proposed model learns a geodesic
with small values of \bTheta, 
approximating a static rank-$k$ subspace
similar to a rank-$k$ SVD.
For comparison, Figure~\ref{fig:syn_loss}
also shows the data error,
which is the norm of the residual between the projected noisy data and the noiseless data.
These plots show that rank-$2k$ SVD
overfits the noise and has a higher error.
The proposed method has a lower error than rank-$2k$ SVD
for any assumed rank greater than or equal to
the true rank on geodesic data.

\new{Figure~\ref{fig:streaming} compares the proposed
geodesic algorithm to several online subspace tracking algorithms,
including 
GROUSE~\citep{balzano2010online}, Oja's algorithm \cite{oja1982simplified},
ISVD~\citep{bunch1978updating},
and PETRELS~\citep{chi2013petrels}. We used $d=20, k=2, \ell=1$ and two noise levels. 
The proposed method consistently
shows better performance (lower error) on geodesic data. See \ref{apx:exp} for detailed information.  
}
Appendix~\ref{apx:exp} also contains 
additional synthetic experiments
exploring a 2D loss surface
and the effect of rank on the rate of convergence.

\begin{figure*}
  \centering
   \includegraphics[width=\linewidth]{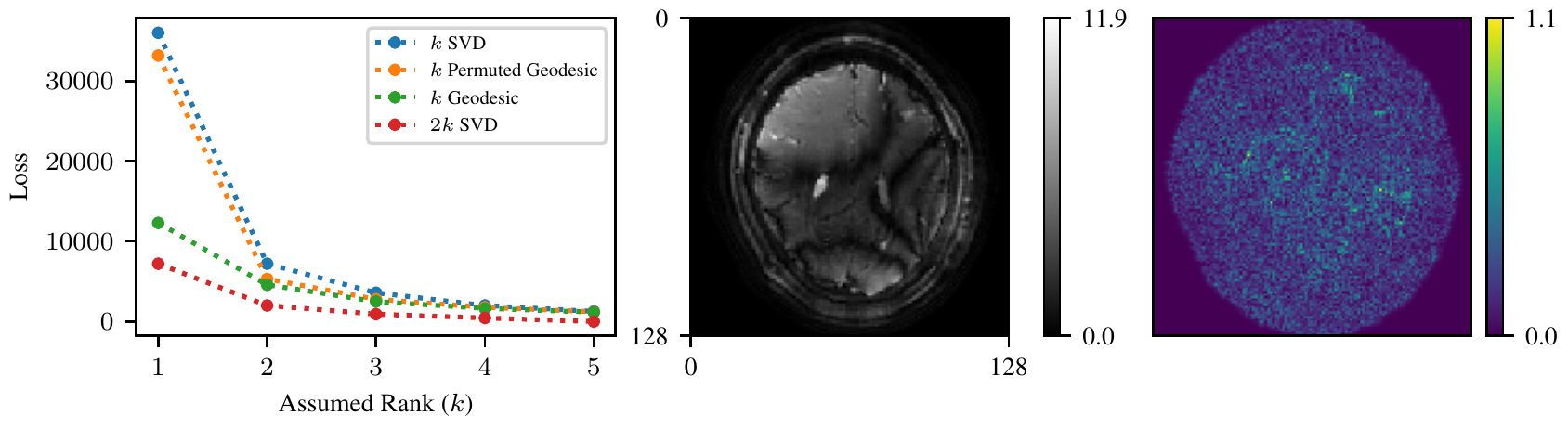}
  \caption{{\bf(Left)} Loss for OSSI dynamic fMRI data.
  In the \(k=1\) case,
  the proposed method has a loss close to rank-$2k$ SVD
  and permuting the data produces a loss closer to rank-$k$ SVD,
  which is similar to Figure~\ref{fig:syn_loss} (top, left).
  From this behavior,
  we infer that the data likely has rank-$1$ geodesic structure.
  {\bf(Center)} Magnitude of reconstructed OSSI image at a single slow and fast test time point.
  For each pixel, fast time points were collected in a vector and 
  a rank-$1$ geodesic was fit across odd slow time points for training.
  Even slow time points were then projected onto the geodesic for test.
  {\bf(Right)} The magnitude difference map of the reconstruction (center) against the true test point.
  The geodesic model appears to have smoothed the image, possibly removing noise.
  The NRMSD across the test points was 0.10.
  This is slightly larger than using a rank-$2$ SVD across the batched data which has a NRMSD of 0.08.
  While imposing a strict temporal structure,
  we fit nearly as well
  as an unconstrained subspace estimation.
  Figure~\ref{fig:ossi_recon} shows
  ground truth and SVD reconstructions. }\label{fig:ossi_loss}
\end{figure*}

\subsection{fMRI Data}

We show the effectiveness of the proposed method
by applying it to (fully anonymized) dynamic functional MRI data
collected with institutional review board approval.
An effective data model for fMRI
could be applied as part of 
an advanced image reconstruction algorithm, 
allowing for reduced scan times and 
higher temporal resolution 
without sacrificing image quality.
We leave a full investigation of
joint reconstruction and modeling
as future work
and, here, show only the viability
of this model on fMRI data.
In particular, we apply the proposed method
on data collected with an oscillating steady state imaging (OSSI)~\citep{guo:20:oss}
acquisition on a 3T GE MR750 scanner.
Appendix~\ref{apx:ossi} 
and
\cite{guo:20:oss}
provide
more details on acquisition and reconstruction
parameters and example data.
The OSSI acquisition rapidly cycles through 10 (\(t_{\mathrm{fast}}\))
different acquisition settings
and then repeats this (\(t_{\mathrm{slow}}\))
for the duration of the scan.
During the scan,
subject breathing and scanner drift
lead to slowly varying subspace changes
that we hypothesized
are suitable for a geodesic model.
The scanner drift is approximately linear in time,
so equally spaced $t_i$ values seems reasonable.
The measurements are dynamic, high dimensional,
and show redundant anatomical structure.
As is common for image subspace models,
we model a spatial patch of data.

Figure~\ref{fig:ossi_loss} shows the loss of applying
the proposed method for a variety of ranks.
Similar to Figure~\ref{fig:syn_loss},
we show a comparison to rank-$k$ and rank-$2k$ SVD
and the proposed method on the data permuted.
From this figure,
we can see that the OSSI data appears to be 
well modeled by a rank 1 geodesic;
the proposed method with rank 1 performs similarly to
rank-$k$ SVD and permuting the data significantly
increases the loss to that of rank-$k$ SVD.


\subsection{Video denoising}\label{subsec:denoising}
\begin{figure*}
    \begin{subfigure}[b]{0.31\textwidth}
         \centering
        \includegraphics[width=\textwidth]{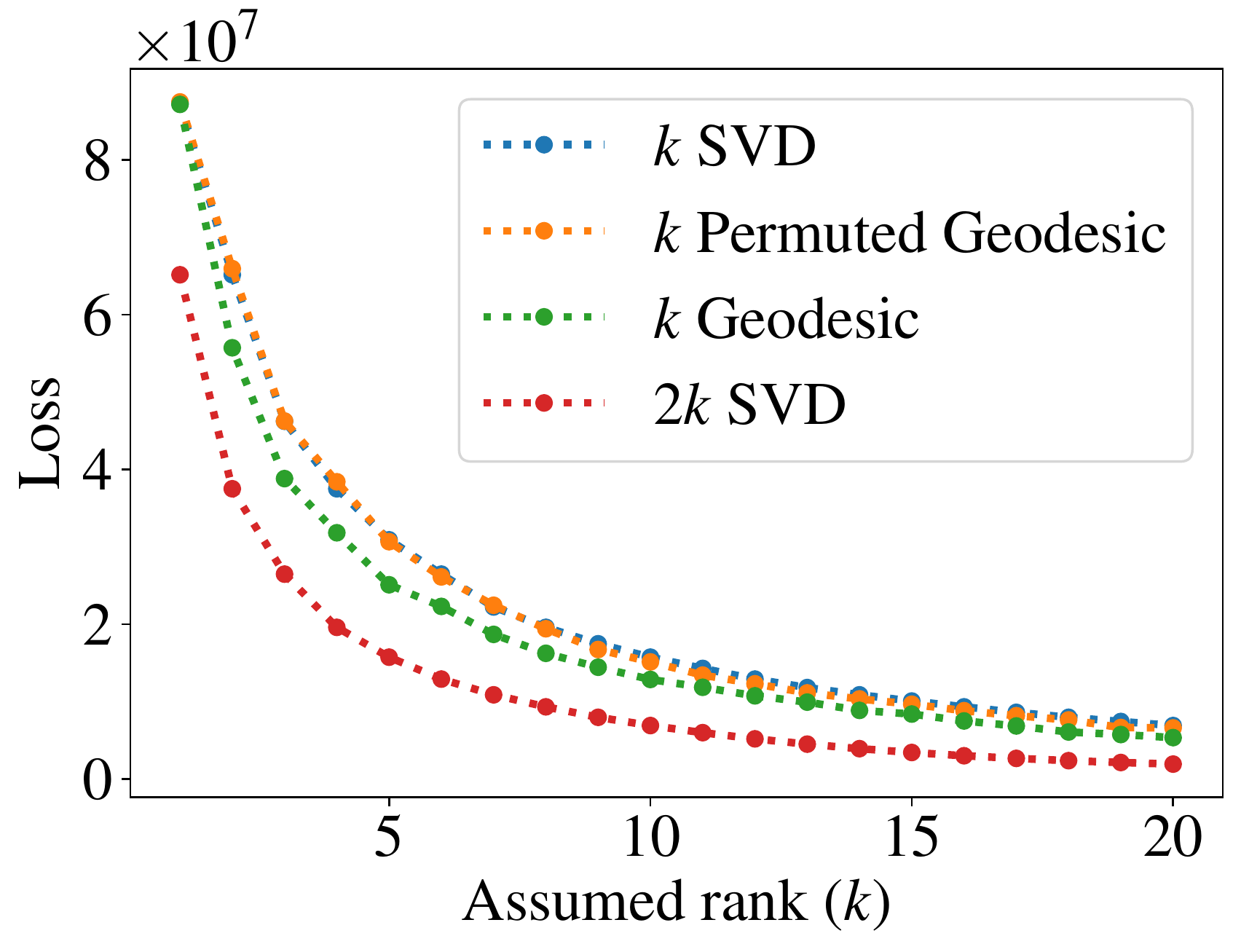}
        \caption{}
        \label{fig:video_recon}
     \end{subfigure}
     \hfill
     \begin{subfigure}[b]{0.31\textwidth}
         \centering
        \includegraphics[width=\textwidth]{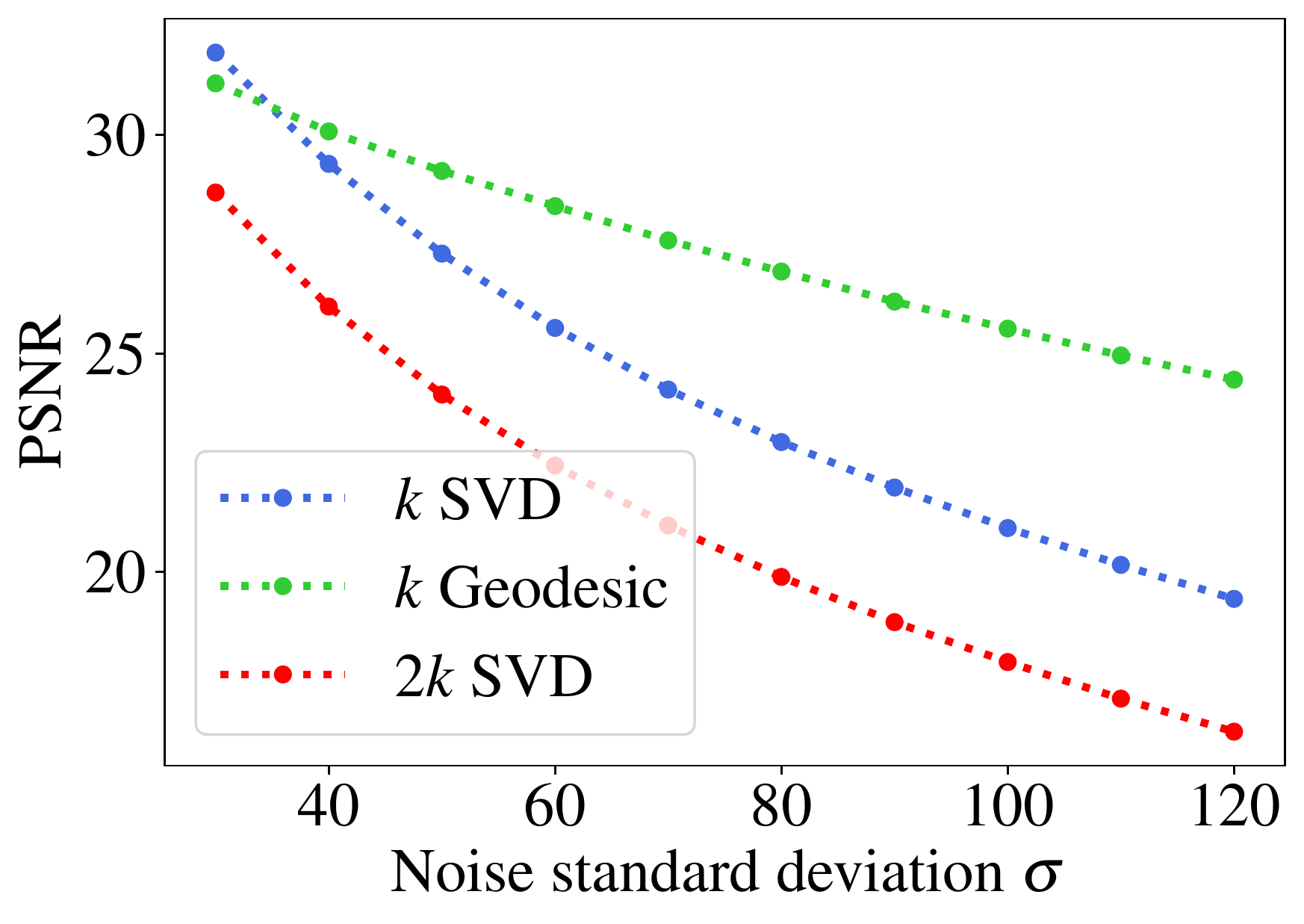}
        \caption{}
        \label{fig:frame125}
     \end{subfigure}
      \hfill
    \begin{subfigure}[b]{0.36\textwidth}
         \centering
        \includegraphics[width=\textwidth]{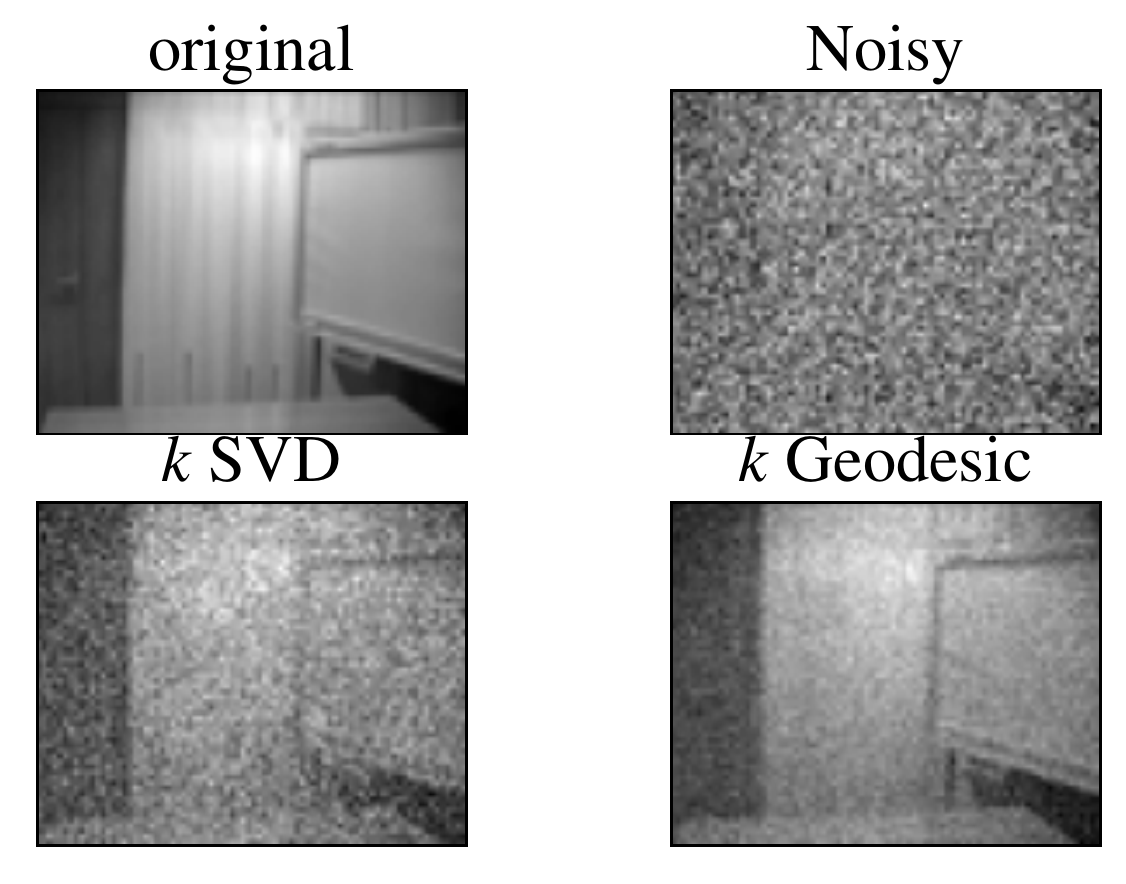}
        \caption{}
        \label{fig:videos}
     \end{subfigure}
     \caption{Quantitative evaluation of geodesic subspace model for video data. In (a) loss from \eqref{eqn:loss2} is plotted for a video sequence containing 260 frames/images. Loss is plotted against different values of assumed rank of data $k$. In (b) we added AWGN to the video data and then apply rank-$k$ SVD, rank-$2k$ SVD, and the geodesic model to denoise the noisy version of video with $k=10$ and $\ell=4$. (c) Visual example of denoising frame 125 in the Curtain video sequence with AWGN of $\sigma=110$. The geodesic model was able to denoise the noisy image more effectively than SVD.}
     \label{fig:video_explore}
\end{figure*}


In this section we apply the geodesic data model for a video denoising application.
The video sequence used in this experiment is the first 260 frames of the Curtain video dataset \citep[Section V-A]{li2004statistical}.
The video sequence has variations
due to a moving curtain and a person entering the scene at the end of the video. See more details and results for other videos in Appendix \ref{apx:vid}.

We start by showing that the geodesic model is a good choice for this video data. Along similar lines as the fMRI data, we achieve this goal by computing loss (representation error) for approximating the video using rank-$k$ SVD, rank-$2k$ SVD, the proposed geodesic method, and applying the geodesic method after reordering/permuting the frames in video sequence.
Again, the rationale behind permuting the frames is that if there is no temporal correlation in the frames
then permuting the data would not have any negative impact on the loss.
Figure~\ref{fig:video_explore}
shows the training loss 
as a function of the rank $k$
and the PSNR of the denoised video
as a function of added noise level.
The training loss for the geodesic model lies
in between $k$ and $2k$ cases,
similar to the simulated data.
More importantly, applying the geodesic model to permuted data
degrades the loss,
confirming that there is temporal correlation for the geodesic model to exploit.
We add additive white Gaussian noise (AWGN)
with different values of standard deviation $\sigma$ to the video sequence
and applied rank-$k$ SVD, rank-$2k$ SVD, and the proposed geodesic subspace model
to denoise the video sequence.
The quality of the denoised image is measured using the peak signal to noise ratio (PSNR),
defined as:

\vspace{-5mm}
\begin{align*}
    \text{PSNR} = 20\log_{10}\left( \frac{255}{\frac{1}{\sqrt{d\times\ell}} \normFr{\X_i -\hat\X_i}} \right).
\end{align*}

\vspace{-3mm}
Figure~\ref{fig:video_explore} shows that
rank-$2k$ SVD has the worst denoising performance
(due to overfitting the very large amount of noise)
while the proposed geodesic method has the best performance in the noisiest regime. 
For visual evidence,
Figure~\ref{fig:videos} illustrates the denoising performance
of rank-$k$ SVD and geodesic model
for frame 125 in the sequence
when noise of $\sigma=110$ is added.
Again, similar to the simulated data, the superior denoising performance of geodesic model is quite evident.

\section{Conclusion, Discussion, and Future Work}
\vspace{-2mm}

This work
proposed a model and algorithm
for dynamic subspace estimation,
or batch-computed subspace tracking.
The proposed method is sample efficient
and the optimization requires no hyperparameters
beyond the assumed rank of the data.
The model and method are applicable to real data,
as shown on dynamic fMRI data and video data,
and the single geodesic we studied here represents a major building block
for a very general piecewise geodesic model.
While we can apply the proposed methods
on chunks of data to learn a series of geodesics,
we leave it as future work to develop a more
efficient way to learn a continuous piecewise geodesic
and its change points.

The proposed method
also requires knowledge of the sample time points \(t_i\).
This is reasonable for temporal data with actual sample times, but there may be problems where we do not know the sample times
or we want to model an unknown varying velocity across the geodesic.
In these cases, we would need a method to estimate suitable \(t_i\)
which we leave for future work.

The time complexity of our algorithm
is dominated by the computation of the \(d \times 2k\) SVD
in the \H, \Y update each iteration.
It is possible that a tighter majorizer
could increase the efficiency of this update.
We leave it as future work to further accelerate the algorithm.

\laura{The 
loss proposed here is nonconvex. 
Major open questions of interest are therefore
understanding the landscape properties of this
objective function, algorithmic convergence to 
stationary points or minimizers, and algorithmic initialization.}
Figure~\ref{fig:phase} exhibits some brighter patches
in the \(T > 2k\) regime
where a few instances appear to have converged to poor local minima.
These instances seem relatively uncommon
for the geodesics considered here.
\laura{We leave it as future work to develop theory for
the loss landscape,
theoretical bounds for geodesic recovery, and 
related initialization techniques.}

Finally, subspace tracking is often applied to problems where data points are modeled as they arrive, e.g.,
in array processing and communications.
It would therefore also be of great interest to develop a streaming algorithm for the piecewise geodesic model.


\section{Acknowledgements}

The authors thank Shouchang Guo and Doug Noll
for sharing the dynamic fMRI data.
This work was supported in part by AFOSR YIP award FA9550-19-1-0026, NSF Grant IIS 1838179,
and NIH Grants U01 EB026977 and R01 EB023618.



\providecommand{\myurl}[1][]      {\url{http://web.eecs.umich.edu/~fessler#1}\xspace} 

\small
\bibliography{bib-jf,bib-other} 


\newpage
\appendix

\begin{center}
{\bf \Large Supplementary Material for \\ ``Dynamic Subspace Estimation with Grassmannian Geodesics''}
\end{center}

\section{Additional Algorithmic Derivations and Details}

\subsection{Derivation of \texorpdfstring{(\H, \Y)}{(H,Y)} 
Update as Majorize Minimize Step}\label{apx:HY_mm}

Let \(\Q \defequ [\H\ \Y]\) and \(\Z_i \defequ [\cos{\bTheta t_i} ; \sin{\bTheta t_i} ]\).
Then our model can be written as
\begin{align}
    \U_i &= \H\cos{\bTheta t_i}+\Y\sin{\bTheta t_i}
    = \Q\Z_i
.\end{align} 
To form a linear majorizer for loss with respect to \Q,
we first derive its unconstrained gradient
\begin{align}
    \loss{\Q} &= -\sum_{i=1}^T \normFr{\X_i^\top\Q\Z_i}^2 + c
    = -\sum_{i=1}^T \trace{\Q^\top\X_i\X_i^\top\Q\Z_i\Z_i^\top} + c
    \\
    \nabla_{\Q}\loss{\Q^|} &= -\sum_{i=1}^T 2\X_i\X_i^\top\Q^|\Z_i\Z_i^\top
.\end{align}

We can form a linear majorizer for the loss
\begin{align}
    g(\Q; \Q^|) &\defequ \trace{\Q^\top\nabla_{\Q}\loss{\Q^|}} + c
    \geq \loss{\Q^|}
.\end{align}
Note that \(g(\Q^|; \Q^|) = \loss{\Q^|}\), and it is linear and continuous.
The above inequality only needs to hold for \(\Q \in \stief^{d \times 2k}\).
It currently holds for \(\reals^{d \times 2k}\)
so there is room for a tighter majorizer.

Following the work of \cite{breloy:21:mmo},
we can write \(g(\Q; \Q^|) = -2\trace{\Q^\top R(\Q^|)} + c\) 
for matrix function \(R(\Q^|)= \sum_{i=1}^T \X_i\X_i^\top\Q^|\Z_i\Z_i^\top\).
We can then minimize the linear majorizer \(g\) with a Stiefel manifold constraint
simply by projecting its negative gradient \(R(\Q^|)\) onto the Stiefel manifold.
The update is then given by
\begin{align}
    \hat\Q &= \argmin{\Q\in\stief^{d\times 2k}}
    \normF{R(\Q^|) - \Q}^2 \\
    &= \argmin{\Q\in\stief^{d\times 2k}}
    \normF{\left(\sum_{i=1}^T \X_i\X_i^\top\Q^|\Z_i\Z_i^\top\right) - \Q}^2 \\
            &= \W\V^\top, \label{eqn:lin_maj}\\
            &\text{  where  } 
            \W\bSigma\V^\top = \sum_{i=1}^T \X_i\X_i^\top\Q^|\Z_i\Z_i^\top\nonumber\\
            &\hspace{60pt}= \sum_{i=1}^T \left[\X_i\hat{\G_i}^\top\cos{\bTheta t_i}\quad
                                               \X_i\hat{\G_i}^\top\sin{\bTheta t_i}\right]\nonumber
,\end{align}
where in the last line we have let \(\hat{\G_i}^\top = \X_i^\top\U_i\iterk = \X_i^\top\Q^|\Z_i\).

We can derive this same update as a block coordinate update on
\Q and \set{\G_i}_{i=1}^T.
We start with our loss \eqnref{eqn:loss2} 
without projecting out \set{\G_i}_{i=1}^T and 
substitute our geodesic model \eqnref{eqn:model}.
The \((\H,\Y)\) update with 
fixed \bTheta and \(\set{\G_i}_{i=1}^T\) 
can be minimized
by recognizing it as a generalized Procrustes problem
\begin{align}
    \hat\H, \hat\Y &=
    \argmin{\H, \, \Y\in\stief^{d\times k}, \, \H^\top\Y=\mathbf{0}} 
          \sum_{i=1}^T \normFr{\X_i - \left(\H\cos{\bTheta t_i} +
                                            \Y\sin{\bTheta t_i}\right)\G_i}^2 \\
    [\hat\H\ \hat\Y] &=
    \argmin{[\H\ \Y]\in\stief^{d\times 2k}} 
          \sum_{i=1}^T \normF{\X_i - [\H\ \Y]\begin{bmatrix}
                                               \cos{\bTheta t_i} \\
                                               \sin{\bTheta t_i}
                                              \end{bmatrix}\G_i}^2 \\
        &= \W\V^\top, \label{eqn:procrustes}\\
            &\text{  where  } 
            \W\bSigma\V^\top = \sum_{i=1}^T 
            \X_i\left(\begin{bmatrix}\cos{\bTheta t_i} \\
                                     \sin{\bTheta t_i}\end{bmatrix}\G_i\right)^\top\nonumber\\
            &\hspace{60pt}= \sum_{i=1}^T \left[\X_i\G_i^\top\cos{\bTheta t_i}\quad
                                               \X_i\G_i^\top\sin{\bTheta t_i}\right]\nonumber
.\end{align}
This Procrustes step involves a single SVD
of the \(d \times 2k\) matrix shown in the last line.
While derived on different losses, Updates \eqnref{eqn:lin_maj} and \eqnref{eqn:procrustes} yeild the same update for \((\H, \Y)\).

\subsection{Derivation of \texorpdfstring{\bTheta}{theta} Update}\label{apx:theta_mm}

We derive a majorize minimize iteration for \bTheta.
First we simplify the loss 
and highlight the separability of the loss 
with respect to the diagonal elements of \bTheta.
We then construct majorizers for each term in the simplified loss
using a translated Huber majorizer.
The update is then given by minimizing the sum of these majorizers.

We note that while we may refer to \bTheta as arc distances,
we do not constrain the elements of \bTheta to be non-negative.
Conceptually, the negative values of \bTheta represent walking
in the opposite direction on the surface of the Grassmannian,
\ie, they are \emph{signed} arc distances.

\subsubsection{Simplifying the Loss}

We start by simplifying the loss
\begin{align}
    \hat\bTheta &= \argmin{\bTheta} \ming{\set{\G_i}} \sum_{i=1}^T 
        \normFr{\X_i - \left(\H\cos{\bTheta t_i}
                            +\Y\sin{\bTheta t_i}\right)\G_i}^2 \\
%
    &= \argmin{\bTheta} - \sum_{i=1}^T 
        \normFr{\X_i^\top\left(\H\cos{\bTheta t_i}+
                           \Y\sin{\bTheta t_i}\right)}^2 \\
    \begin{split}
    &= \begin{aligned}[t] \argmin{\bTheta}- \sum_{i=1}^T 
        \trace{\X_i^\top\left(
                        \H\cos{\bTheta t_i}^2\H^\top 
                        + 2\real{\H\cos{\bTheta t_i}\sin{\bTheta t_i}\Y^\top} \right.\right. \hphantom{XXX}\\ \left.\left.
                        + \Y\sin{\bTheta t_i}^2\Y^\top\right)\X_i}
    \end{aligned}\end{split}\\
    \begin{split}
    &= \begin{aligned}[t] \argmin{\bTheta}- \sum_{i=1}^T 
        \trace{
            \cos{\bTheta t_i}^2\H^\top\X_i\X_i^\top\H \right. \hphantom{XXXXXXXXXXXXXXXXXXXX}\\ \left.
            +2\cos{\bTheta t_i}\sin{\bTheta t_i}\real{\Y^\top\X_i\X_i\H} \right. \hphantom{XXXXX}\\ \left.
            +\sin{\bTheta t_i}^2\Y^\top\X_i\X_i^\top\Y}
    \end{aligned}\end{split}\\
    \begin{split}
    &= \begin{aligned}[t]  \argmin{\bTheta}- \sum_{i=1}^T 
        \sum_{j=1}^k
            \cos{\theta_j t_i}^2{\left[\H^\top\X_i\X_i^\top\H \right]}_{j,j} \hphantom{XXXXXXXXXXXXXXXXXXX}\\ 
            +2\cos{\theta_j t_i}\sin{\theta_j t_i}{\left[\real{\Y^\top\X_i\X_i\H}\right]}_{j,j} \hphantom{XXXX}\\ 
            +\sin{\theta_j t_i}^2{\left[\Y^\top\X_i\X_i^\top\Y\right]}_{j,j}
    .\end{aligned}\end{split}
\end{align}
We solve the problem of optimizing \bTheta
by updating
each of its diagonal elements \theta_j separately.
We define the following constants
\begin{align}
    \alpha_{i,j} &= {[\H^\top\X_i\X_i^\top\H]}_{j,j} \\
    \beta_{i,j} &= \real{{[\Y^\top\X_i\X_i^\top\H]}_{j,j}} \\
    \gamma_{i,j} &= {[\Y^\top\X_i\X_i^\top\Y]}_{j,j}
.\end{align}
Our optimization problem for each $j=1,\ldots,k$ is now
\begin{align}
    \hat\theta_j &= \argmin{\theta_j}- \sum_{i=1}^T 
            \alpha_{i,j}\cos{\theta_j t_i}^2
            +2\beta_{i,j}\cos{\theta_j t_i}\sin{\theta_j t_i}
            +\gamma_{i,j}\sin{\theta_j t_i}^2
.\end{align}
Using the following trigonometric identities
\begin{align}
    &2\cos{x}\sin{x} = \sin{2x} \\
    &\cos{x}^2 + \sin{x}^2 = 1 \\
    &\cos{x}^2 = \frac12\left(\cos{2x} + 1\right) \\
    &a\cos{x} + b\sin{x} = \sqrt{a^2 + b^2}\cos{x - \arctan\!2(b,a)}
,\end{align}
we further simplify the loss:
\begin{align}
    \hat\theta_j &= \argmin{\theta_j}- \sum_{i=1}^T 
            \alpha_{i,j}\cos{\theta_j t_i}^2
            +\beta_{i,j}\sin{2\theta_j t_i}
            +\gamma_{i,j}\sin{\theta_j t_i}^2\\
%
    &= \argmin{\theta_j}- \sum_{i=1}^T 
            (\alpha_{i,j}-\gamma_{i,j})\cos{\theta_j t_i}^2
            +\beta_{i,j}\sin{2\theta_j t_i}
            +\gamma_{i,j}\left(\cos{\theta_j t_i}^2 + \sin{\theta_j t_i}^2\right)\\
%
    &= \argmin{\theta_j}- \sum_{i=1}^T 
            (\alpha_{i,j}-\gamma_{i,j})\frac12\left(\cos{2\theta_j t_i} + 1\right)
            +\beta_{i,j}\sin{2\theta_j t_i}
            +\gamma_{i,j}\\
%
    &= \argmin{\theta_j}- \sum_{i=1}^T 
            r_{i,j}\cos{2\theta_j t_i - \phi_{i,j}} + b_{i,j}
,\end{align}
where
\begin{align}
    r_{i,j} &= \sqrt{ {\left( \frac{\alpha_{i,j}-\gamma_{i,j}}{2} \right)}^2 
                  + \beta_{i,j}^2 } \\
    \phi_{i,j} &= \arctan\!2\left(\beta_{i,j}, \frac{\alpha_{i,j} - \gamma_{i,j}}{2}\right) \\
    b_{i,j} &= \frac{\alpha_{i,j} + \gamma_{i,j}}{2}
.\end{align}

\subsubsection{Constructing a majorizer}\label{apx:theta_mm:maj}

To majorize our loss function,
we construct a quadratic majorizer for each term of the form
\begin{equation}
    \f_{i,j}(\theta_j) \defequ -r_{i,j}\cos(2\theta_j t_i - \phi_{i,j}) + b_{i,j}
.\end{equation}

To be a majorizer at a point \theta_j^|, 
we require 
\(\q_{i,j}{\theta_j^|;\theta_j^|} = \f_{i,j}{\theta_j^|}\) (equal at the point of construction) 
and \(\q_{i,j}{\theta_j;\theta_j^|} \geq \f_{i,j}{\theta_j}\) (greater than or equal to the loss everywhere). 
This can be achieved with a quadratic of the form
\begin{equation}
    \q_{i,j}(\theta_j; \theta_j^|) = \f_{i,j}(\theta_j^|) 
                                + \f_{i,j}^.(\theta_j^|)(\theta_j-\theta_j^|) 
                                + \frac12\w_{\f_{i,j}}(\theta_j^|){(\theta_j-\theta_j^|)}^2 
,\end{equation}
where \(\f_{i,j}^.(\theta_j) = 2r_{i,j}t_i\sin{2\theta_j t_i - \phi_{i,j}}\) is the derivative of \f_{i,j}
and \w_{\f_{i,j}} is an appropriate curvature (or ``weighting'') function.
A simple option is \(\w_{\f_{i,j}}(\theta_j^|) = L_{\f_{i,j}^.}\) the Lipschitz constant of the derivative.
Minimizing the resulting majorizer yields the standard fixed step size gradient descent algorithm.
A tighter majorizer will touch our original function at two or more points.
Note that \(\frac{\phi_{i,j}}{2 t_i}\) is a minimizer 
and our function is symmetric and quasi-convex on the interval 
\(\left[\frac{\phi_{i,j} - \pi}{2 t_i}, \frac{\phi_{i,j} + \pi}{2 t_i} \right]\)
about this point. Our approach will be to construct a curvature function \(\w_{\f_{i,j}}^|\) 
for points in this interval and 
periodically extend it to construct the final curvature function \(\w_{\f_{i,j}}\).
Because \f_{i,j} is symmetric about \(\frac{\phi_{i,j}}{2 t_i}\), 
our majorizer will touch at two points 
when the axis (and minimizer) of \q_{i,j} is \(\frac{\phi_{i,j}}{2 t_i}\),
equivalently when its gradient at this point equals zero
\begin{equation}
    \q_{i,j}^.(\frac{\phi_{i,j}}{2 t_i}; \theta_j^|) = 0
.\end{equation}
Solving for \(\w_{\f_{i,j}}^|(\theta_j^|)\) yields
\begin{equation}
\w_{\f_{i,j}}^|(\theta_j^|) = \frac{\f_{i,j}^.(\theta_j^|)}{\theta_j^| - \frac{\phi_{i,j}}{2t_i}}
,\end{equation}
which can be recognized as a (translated) Huber curvature function.
For the case when \(\theta_j^| = \frac{\phi_{i,j}}{2t_i}\),
we define \(\w_{\f_{i,j}}(\frac{\phi_{i,j}}{2t_i}) = 4 t_i^2 r_{i,j}\),
which is its limit point.

Forming \w_{\f_{i,j}} by periodically extending \w_{\f_{i,j}}^| 
only requires periodically extending the denominator, 
since the numerator is already periodic. 
The resulting periodic version of the curvature function is
\begin{equation}
\w_{\f_{i,j}}(\theta_j^|) = \begin{cases}
                \frac{\f_{i,j}^.(\theta_j^|)}{
                \mod\!\left(
                    (\theta_j^| - \frac{\phi_{i,j}}{2t_i}) + \frac{\pi}{2t_i}, 
                    \frac{2\pi}{2t_i}
                \right) - \frac{\pi}{2t_i} } & \theta_j^| \neq \frac{\phi_{i,j} + 2\pi m}{2t_i}, m\in\ints\\
                4 t_i^2 r_{i,j} & \theta_j^| = \frac{\phi_{i,j} + 2\pi m}{2t_i}, m\in\ints
                .\end{cases}
\end{equation}

\begin{figure}
  \centering
   \includegraphics[width=\linewidth]{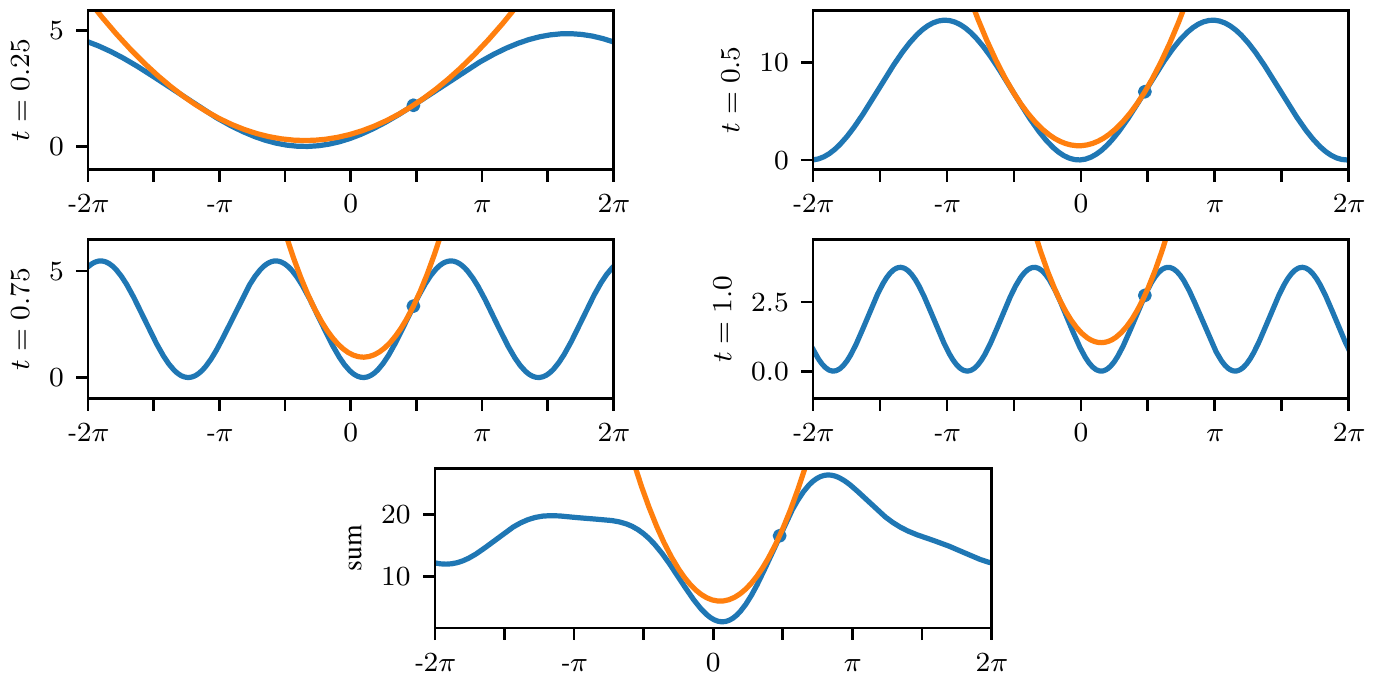}
  \caption{An example of four cosines (top two rows, blue) 
  that sum to form the (nonconvex) loss
  for a single \theta_j (bottom row, blue). 
  For each cosine function, 
  we construct a quadratic majorizer (top two rows, orange)
  at a point \theta_j^| (blue dot). 
  The sum of these individual quadratic majorizers 
  form a quadratic majorizer for the loss (bottom, orange) 
  that has a closed-form minimizer.
  Although the loss is non-convex, 
  distance-minimizing geodesics will have 
  \(\theta_j \in [-\pi/2, \pi/2]\). 
  On this interval, 
  the loss is often well-behaved 
  (here, quasi-convex).
  }\label{fig:theta_maj}
\end{figure}

The final majorizer for our loss function of \theta_j is then
\begin{equation}
    \q_{j}{\theta_j; \theta_j^|} = \sum_{i=1}^T \q_{i,j}{\theta_j; \theta_j^|}
.\end{equation}
Figure~\ref{fig:theta_maj} shows an example loss and the constructed majorizer.

\subsubsection{Minimizing the Majorizer}

Because the majorizer is a sum of one-dimensional quadratics,
we minimize it by setting its derivative to zero and solving for \theta_j
\begin{align}
     \q_{j}^.{\theta_j; \theta_j^|} &= \sum_{i=1}^T \q_{i,j}^.{\theta_j; \theta_j^|} \\
     0 &= \sum_{i=1}^T \f_{i,j}^.(\theta_j^|) + \w_{\f_{i,j}}(\theta_j^|) (\theta_j-\theta_j^|)
     = \left(\sum_{i=1}^T \f_{i,j}^.(\theta_j^|)\right) + \left(\sum_{i=1}^T \w_{\f_{i,j}}(\theta_j^|) \right)(\theta_j-\theta_j^|)\\
     \theta_j &= \theta_j^| - \frac{\sum_{i=1}^T \f_{i,j}^.(\theta_j^|)}{\sum_{i=1}^T \w_{\f_{i,j}}(\theta_j^|) }
.\end{align}

Iteratively constructing majorizers and minimizing them yields the following descent scheme
\begin{equation}{
    \theta_j\iterk[+1] = \theta_j\iterk - 
                        \frac{\sum_{i=1}^T \f_{i,j}^.(\theta_j\iterk) }
                             {\sum_{i=1}^T \w_{\f_{i,j}}(\theta_j\iterk)}
.}\end{equation}


\begin{figure}[pt!]
  \centering
\begin{algorithm}[H]
\caption{Geodesic Subspace Estimation} 
\label{alg:gse}
\begin{algorithmic}
\REQUIRE \(\set{\X_i, t_i}_{i=1}^T\), $\H\iterz, \Y\iterz, \bTheta\iterz$,
$N=$ \# of outer iterations,
$M=$ \# of inner MM iterations


\FOR{\(\iteridx = 1,\ldots,N\)}
    \STATE \textit{\# \(\H, \Y\) update}

    \STATE \(\displaystyle \U_i\iterk[-1] = 
                \H\iterk[-1]\cos{\bTheta\iterk[-1] t_i} + 
                \Y\iterk[-1]\sin{\bTheta\iterk[-1] t_i} \quad\forall i\)
    \STATE \(\displaystyle \G_i = (\U_i\iterk[-1])^\top \X_i \quad \forall i \)
    \STATE \(\displaystyle \M = \sum_{i=1}^T \left[\X_i\G_i^\top\cos{\bTheta\iterk[-1] t_i}\quad
                                               \X_i\G_i^\top\sin{\bTheta\iterk[-1] t_i}\right]\)
    \STATE \(\displaystyle \W, \bSigma, \V^\top = \mathrm{SVD}(\M)\)
    \STATE \(\displaystyle [\H\iterk\ \Y\iterk] = \W\V^\top\)

    \vspace{1.3ex}
    \STATE \textit{\# \(\bTheta\) update}

    \STATE \(\bTheta\iterk[,0] = \bTheta\iterk[-1]\)
    \FOR{\(j = 1,\ldots,k\)}
    \STATE \(\displaystyle \alpha_{i,j} = {[\H^\top\X_i\X_i^\top\H]}_{j,j} \quad \forall i \)
    \STATE \(\displaystyle \beta_{i,j} = \real{{[\Y^\top\X_i\X_i^\top\H]}_{j,j}} \quad \forall i \)
    \STATE \(\displaystyle \gamma_{i,j} = {[\Y^\top\X_i\X_i^\top\Y]}_{j,j} \quad \forall i \)

    \STATE \(\displaystyle \phi_{i,j} = \arctan\!2\!\left(\beta_{i,j},\frac{\alpha_{i,j}-\gamma_{i,j}}{2}\right) \quad \forall i \)
    \STATE \(\displaystyle r_{i,j} = \sqrt{ {\left( \frac{\alpha_{i,j}-\gamma_{i,j}}{2} \right)}^2 
                  + \beta_{i,j}^2 } \quad \forall i \)

        \FOR{\(m = 1,\ldots,M\)}
            \STATE \(\displaystyle z_j = \sumno_{i=1}^T 2r_{i,j}t_i \sin{2\theta\iterk[,m-1]_j t_i - \phi_{i,j}}\)
            \STATE \(\displaystyle w_j = \sum_{i=1}^T \frac{
                   2 r_{i,j} t_i \sin{2\theta\iterk[,m-1]_j t_i - \phi_{i,j}}}{
                   \mod\!\left(
                        (\theta\iterk[,m-1]_j - \frac{\phi_{i,j}}{2t_i}) + \frac{\pi}{2t_i}, 
                        \frac{2\pi}{2t_i}
                   \right) - \frac{\pi}{2t_i} }\)
            \STATE \(\displaystyle \theta\iterk[,m]_j = \theta\iterk[,m-1]_j - \frac{z_j}{w_j}\)
        \ENDFOR
    \ENDFOR
    \STATE \(\bTheta\iterk = \bTheta\iterk[,M]\)
\ENDFOR

\end{algorithmic}
\end{algorithm}
\end{figure}

\section{Additional Experiments and Details}

\subsection{Synthetic Experiments - Details and Further Investigation} \label{apx:exp}

\paragraph{Rank-1 subspace in $\reals^2$} To provide some simple intuition for our problem,
we present the algorithm applied to learning a rank-1 subspace
in two dimensions.
In this special case,
\bTheta is a scalar \theta,
and \H and \Y can be parameterized
by a single scalar rotation \omega
(up to a sign flip in \Y, which we can absorb into \theta).
Then \(\H = [\cos{\omega} ; \sin{\omega} ]\)
and \(\Y = [-\sin{\omega} ; \cos{\omega} ]\).
Our loss \eqnref{eqn:loss2} then simplifies to a two-dimensional function of \theta and \omega
\begin{equation}
    \loss{\U} = -\sum_{i=1}^T r_{i,1}\cos(2\theta t_i - \phi_{i,1} + 2\omega) + b_{i,1} + c
,\end{equation}
where \(r_{i,1}, \phi_{i,1}, b_{i,1}\) are defined as previously, but with $[\H \, \Y] = \I$
and $c$ is the same constant from \eqnref{eqn:loss2}.

\begin{figure}
  \centering
   \includegraphics[width=\linewidth]{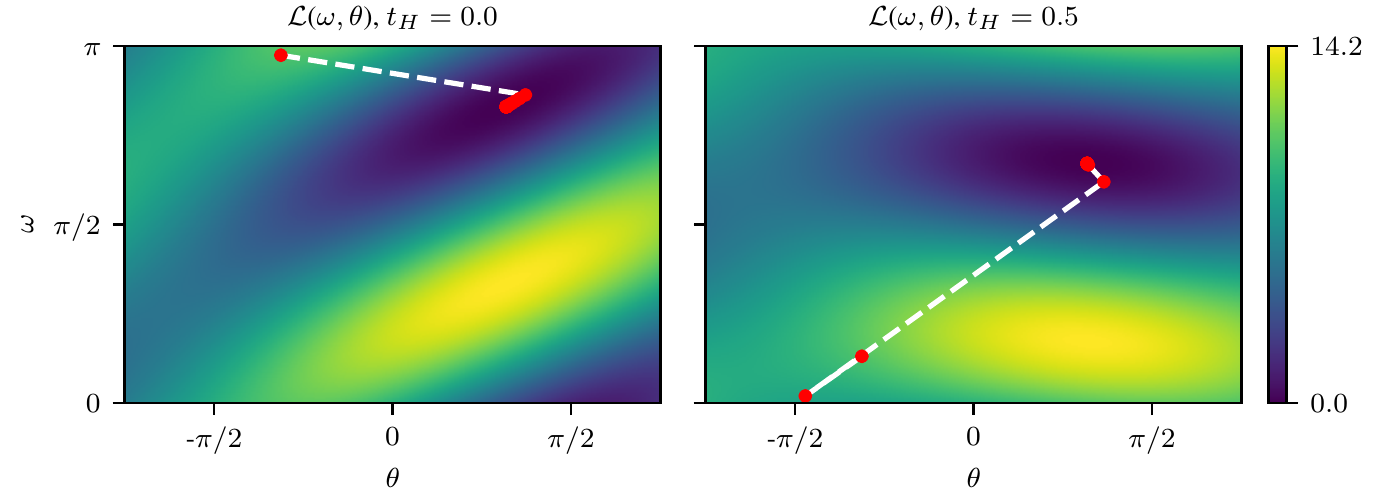}
  \caption{The loss function for learning a rank-1 geodesic in 2D,
    where we have made the transformation
    \(\H = [\cos{\omega} ; \sin{\omega} ]\)
    and \(\Y = [-\sin{\omega} ; \cos{\omega} ]\).
    Note that the y-axes are $\pi$-periodic.
    (Left) The unaltered loss function and associated
    algorithm iterates (red).
    The iterates approach the minimum slowly,
    as the structure of the loss is not well aligned
    with the coordinate directions.
    (Right) The loss after making the transformation
    \(\tilde{t}_i = t_i - t_{\H}\) for \(t_{\H}=0.5\).
    The iterates fully converge in only a few iterations.
    Both sets of iterations are initialized at equivalent points.
  }\label{fig:2dloss}
\end{figure}

Figure~\ref{fig:2dloss} (left) shows this loss function on some noisy synthetic data
with iterates of the proposed algorithm shown in red.
Many iterations take small steps as the minimizers
of \theta and \omega are very interdependent.

Intuitively, this behavior may be because the optimal starting subspace \H
is very dependent on the arc length \theta of the geodesic,
as the method will naturally want to center the geodesic to minimize error.
If we instead parameterized our geodesic by its center subspace,
the resulting \H minimizer would hopefully be more independent of arc length.
This can be done by applying the proposed method after first
transforming the time points by letting \(\tilde{t}_i = t_i - t_{\H}\),
where \(t_{\H} \in [0,1]\) is the point along the geodesic equal to \H.
Figure~\ref{fig:2dloss} (right) shows the loss and associated iterates 
of setting \(t_{\H} = \nicefrac12\).
The proposed algorithm converges in only a few iterations.

\begin{figure}
  \centering
   \includegraphics[width=\linewidth]{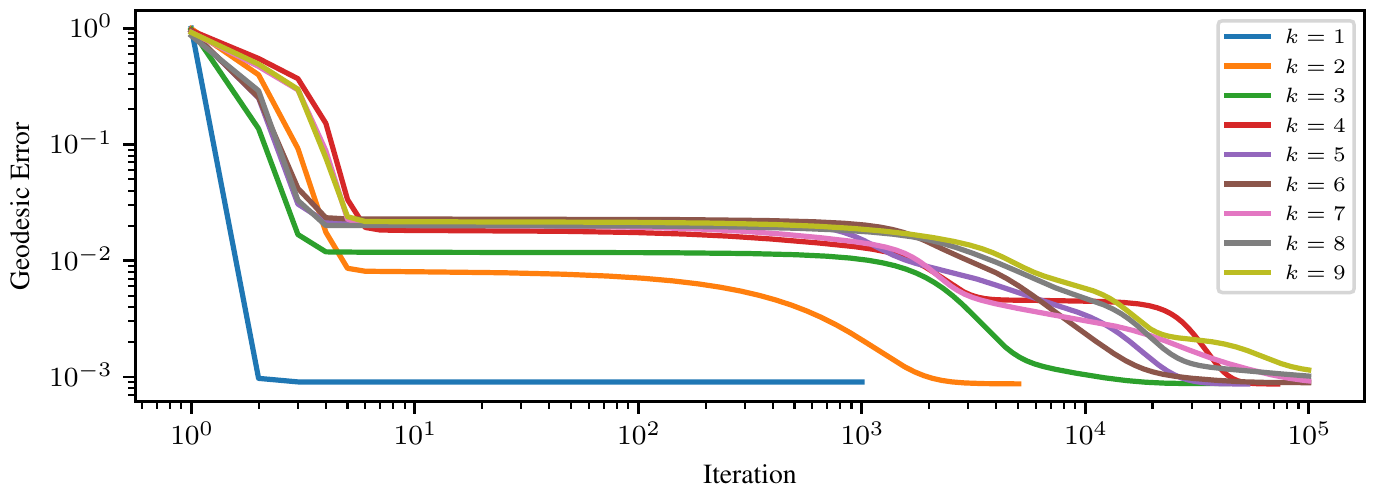}
  \caption{Convergence of the proposed algorithm in geodesic error to the true geodesic for nine planted geodesic models with varying ranks in \reals^{40}. The planted models were used to generate 100 sample points (\(T=100, \ell=1\)) and AWGN with standard deviation \(\sigma = 10^{-3}\) was added. The algorithm was initialized with a random geodesic.
  }\label{fig:syn_conv}
\end{figure}

\paragraph{Comparison to streaming subspace estimation methods}

Here we give more details about Figure~\ref{fig:streaming}, which compares the proposed
geodesic algorithm to several online subspace tracking algorithms,
including 
GROUSE~\citep{balzano2010online}, Oja's algorithm \cite{oja1982simplified,huang2021streaming}, 
Incremental SVD (ISVD)~\citep{bunch1978updating,balzano2018streaming},
and PETRELS~\citep{chi2013petrels}. We used the code for these algorithms provided by the authors of \cite{balzano2018streaming}. We used $d=20, k=2, \ell=1$ and two noise levels, $\sigma = 10^{-1}, 10^{-5}$. 
The proposed method consistently
shows better performance (lower error) on geodesic data.

Each of these comparison methods performs subspace updates in a streaming way, one vector at a time, and so their subspace estimate is time-varying. However, without an explicit time-varying model, there is no natural way to pass over the data multiple times and improve the estimate for each time point, and so this comparison only uses one pass over the data for the streaming algorithms. 

 For Oja's algorithm, we used step-size 0.1, and we can see it requires several vectors before it begins to converge. For the GROUSE algorithm, we used the ``greedy step-size" that is default in the implementation provided for \cite{balzano2018streaming}. For this step-size choice, GROUSE performs quite poorly with high noise but converges somewhat close to the true subspace with a small amount of noise. We note that Oja's algorithm and GROUSE have been shown to be equivalent \cite{balzano2022equivalence} for certain step-sizes, so these two lines are essentially showing behavior for different choices of step-size on the same algorithm. 

The ISVD algorithm performs the incremental singular value decomposition with truncation of the smallest singular values at every step. It therefore quickly estimates the subspace at early iterations, but then drifts from the true subspace since it is not adapting to the new data. The PETRELS algorithm, for which we used a forgetting factor of $0.8$, finds a very nice balance between quick estimation of the early subspace followed by adaptation.

\paragraph{Empirical Error Convergence} Figure~\ref{fig:syn_conv} shows 
the geodesic error per iteration of the proposed algorithm applied to
several synthetically generated planted models.
Generally, the proposed algorithm for rank-$1$ geodesics
converges in only a few iterations,
while larger \(k\) requires an increasing number of iterations to converge.
These experiments were initialized randomly,
but were still able to recover the true geodesic with error at the level of the additive noise.
We occasionally see the algorithm converge to poor local minima 
and fail to recover the true geodesic.
We leave it a future work to make the algorithm more robust to these instances
and to provide theoretical bounds on geodesic recovery.

\subsection{OSSI Dynamic fMRI Dataset Details}\label{apx:ossi}

\begin{figure}
  \centering
   \includegraphics[width=\linewidth]{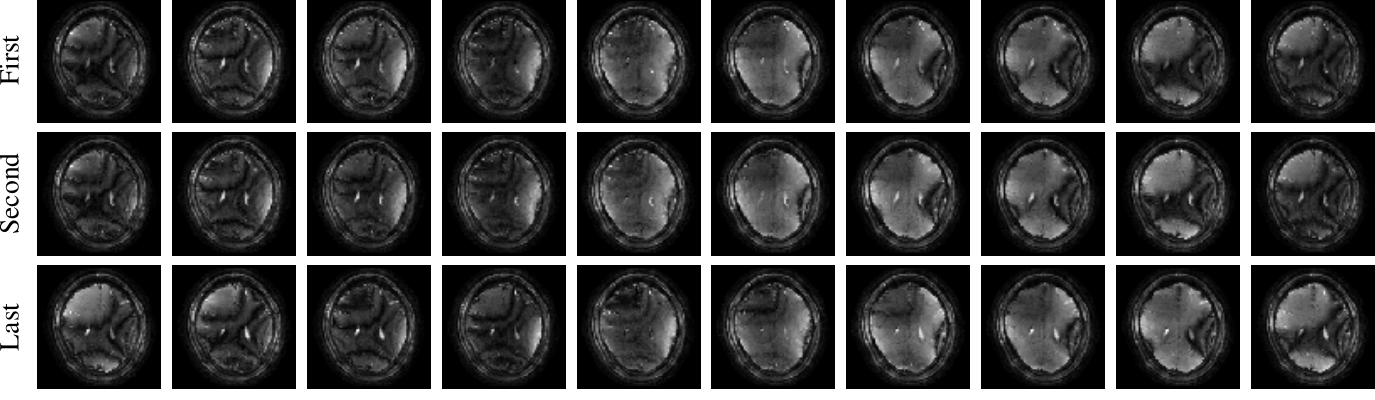}
  \caption{A sample of OSSI acquisitions.
  Each column, referred to as fast time, represents a certain set of acquisition parameters.
  Each row, referred to as slow time, is a complete cycle through these acquisitions, which is done many times.
  Here we only show the first, second, and last slow time set. We see that there is little difference between neighboring slow time points, but that over the course of the scan they change more significantly.
  }\label{fig:ossi_dataset}
\end{figure}

The OSSI dynamic fMRI dataset was acquired
on a 3T GE MR750 scanner with a 32-channel head coil
and is comprised of 167 slow time, 10 fast time and 128 $\times$ 128 spatial samples.
The complex data was fully sampled with a variable-density spiral trajectory
with \(n_i=8\) interleaves, a densely sampled core, and spiral-out readouts.
Detailed OSSI acquisition parameters (TR, $n_c$, TE, and flip angle)
can be found in Guo and Noll~\citep{guo:20:oss}.
The volunteer was given a left versus right 
reversing-checkerboard visual stimulus 
(20~s Left / 20~s Right $\times$ 5 cycles) for 200~s in total.
For reconstruction, the k-space data was compressed to 16 virtual coils
and
ESPIRiT SENSE maps were generated using the BART toolbox~\citep{uecker:14:eae}.
Finally, the images were reconstructed using
conjugate gradient SENSE with a Huber potential via the MIRT toolbox~\citep{fessler:16:irt}.
Figure~\ref{fig:ossi_dataset} shows the magnitude of a sample of the reconstructed images.

\begin{figure}
  \centering
   \includegraphics[width=\linewidth]{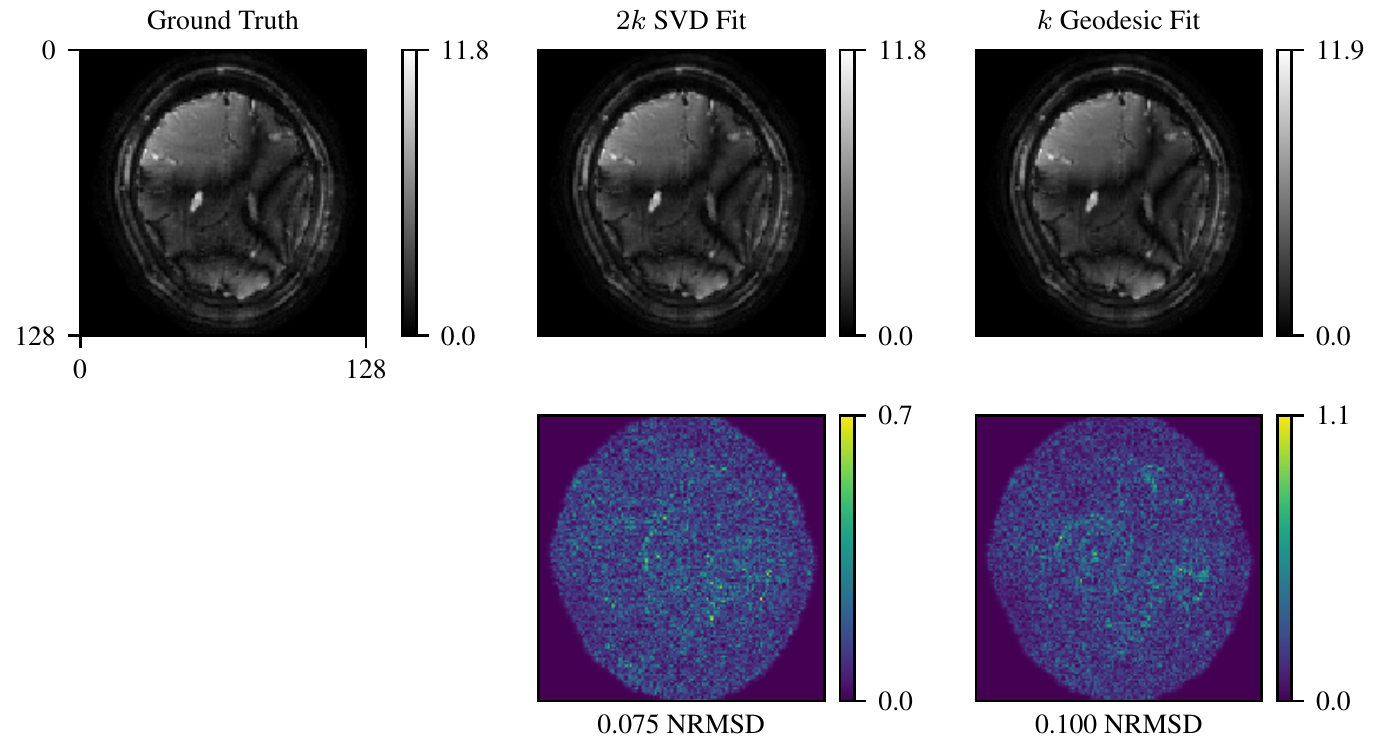}
  \caption{Comparison of OSSI magnitude images reconstructed with a static $2k$ subspace \vs the geodesic model for a particular slow and fast time point (See Figure~\ref{fig:ossi_loss}). Despite being more temporally constrained, the proposed method fits nearly as well.
  }\label{fig:ossi_recon}
\end{figure}

\subsubsection{Additional Figures for OSSI data}
\label{apx:addlossi}

Figure~\ref{fig:ossi_recon} provides comparison to ground truth and rank-$2k$ SVD for the geodesic OSSI reconstruction presented in Figure~\ref{fig:ossi_loss} on a single slow and fast time point.
The proposed geodesic model fits nearly as well despite being a more constrained model.

\begin{figure}
  \centering
   \includegraphics[width=\linewidth]{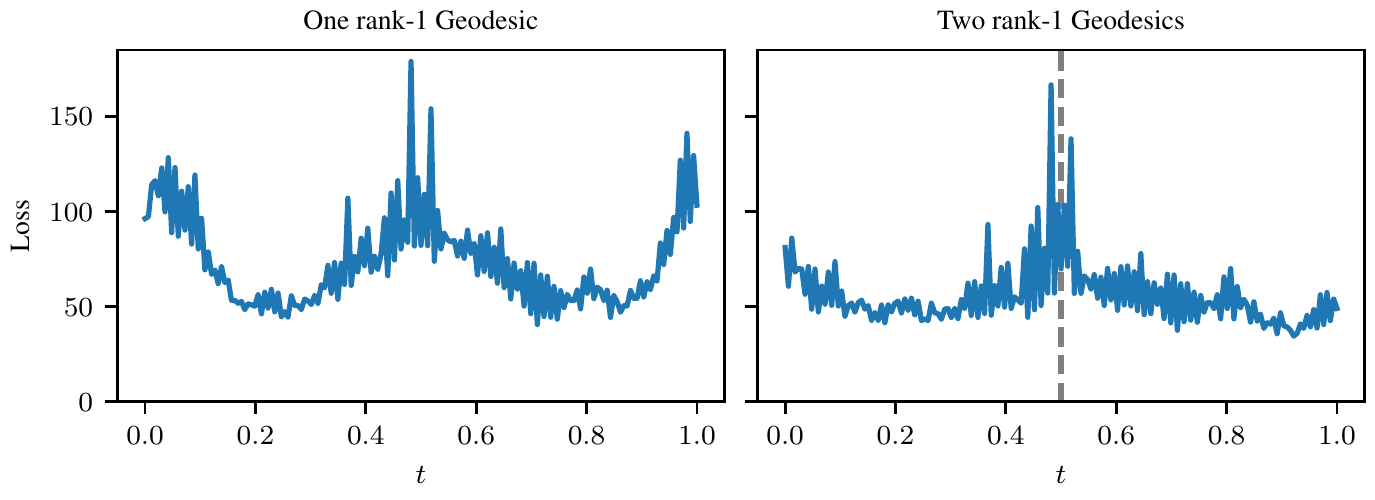}
  \caption{{\bf(Left)} The loss broken down for each time point $t$ for fitting a rank-1 geodesic on OSSI fMRI data. In addition to a seemingly noisy component, the loss also exhibits systematic variations. We hypothesize that these systematic variations result from the limitations of a single geodesic to fit a curve. {\bf(Right)} Instead of fitting a single rank-1 geodesic, we model the data as a piecewise geodesic by fitting a rank-1 geodesic on each half of the data. The resulting loss retains the noisy variations from the single geodesic model, but removes most of the systematic variation.
  }\label{fig:ossi_loss_v_time}
\end{figure}

\paragraph{Piecewise Geodesic} While the OSSI data can be modeled well by a rank-$1$ geodesic,
Figure~\ref{fig:ossi_loss_v_time} provides some initial evidence 
that a piecewise rank-$1$ geodesic may be a better model. In this figure, we first modeled it with a single rank-1 geodesic (as in the main body of the paper), and plotted value of the loss function at each data point $\X_i$ (on the left). From this we can see that the loss varies in a W shape across the data points, which may indicate that a piecewise geodesic is more appropriate. Then we tried two rank-1 geodesics, fitting the first half and second half of the data to separate rank-1 geodesics. When we plot the loss for this model, the shape of the loss is more flat. A more detailed investigation of methods for identifying knots in a piecewise geodesic model is of great interest for future work.

Even when we know the knots
of a piecewise geodesic model,
applying the described geodesic estimation algorithm
on each segment individually
is not guaranteed to
produce a continuous piecewise geodesic.
If we let \(\U_j(t)\) represent the \jth geodesic segment,
then we would like
\(\Span{\U_j(1)} = \Span{\U_{j+1}(0)}\),
or equivalently
\(\U_j(1)\U_j(1)^\top = \U_{j+1}(0)\U_{j+1}(0)^\top,\ \forall j=1\ldots J-1\).
Combining this constraint with our loss
yields the following optimization problem for piecewise geodesics
\begin{align}
    \min{\U_j}& -\sum_{j=1}^J\sum_{i=1}^T\normF{\X_{j,i}^\top\U_j(t_i)}^2 \\
    &\suchthat \U_j(1)\U_j(1)^\top = \U_{j+1}(0)\U_{j+1}(0)^\top\quad
    \forall j=1\ldots J-1 \nonumber
.\end{align}

We can simplify this problem by relaxing the constraint to a penalty.
If we optimize the loss using a BCD approach on each segment,
then the penalized loss for the \jth segment is
\begin{align}
    \min{\U_j} -\sum_{i=1}^T\normF{\X_i^\top\U_j(t_i)}^2
    &+ \lambda \normF{\U_j(0)\U_j(0)^\top-\U_{j-1}(1)\U_{j-1}(1)^\top}^2 \nonumber\\
    &+ \lambda \normF{\U_j(1)\U_j(1)^\top-\U_{j+1}(0)\U_{j+1}(0)^\top}^2 \\
    \min{\U_j} -\sum_{i=1}^T\normF{\X_i^\top\U_j(t_i)}^2
    &- \lambda \normF{\U_{j-1}(1)^\top\U_j(0)}^2
     - \lambda \normF{\U_{j+1}(0)^\top\U_j(1)}^2
,\end{align}
where \X_i is the data
on just the \jth segment.
Intuitively, this loss can be minimized
using the same updates described previously,
but with extra data
\(\X_1^~\defequ\sqrt{\lambda}\U_{j-1}(1)\) and
\(\X_T^~\defequ\sqrt{\lambda}\U_{j+1}(0)\)
at each end of the geodesic,
\(t_1=0\) and \(t_T=1\) respectively.

This formulation will perform better
on a continuous piecewise geodesic
than just applying the proposed method
to each geodesic individually,
but this approach will only approximate
a continuous piecewise geodesic
when \lambda is sufficiently large
(though too large of \lambda
will cause the BCD algorithm
to disregard the data and stall suboptimally).
An effective approach may be to start with
\(\lambda = 0\)
and slowly increase it as the algorithm converges.
This approach may be sufficient
since we are only approximating
real data as being derived from a piecewise geodesic
and still are able to leverage the extra data.
In a large data regime,
enforcing the constraint too strictly
may only lead to larger modeling error.
Thus, a penalized formulation allows us
to balance the fit to the model and the data.
Alternatively,
an ADMM~\citep{boyd:10:doa}
approach could be developed
to satisfy the constraint strictly
in a similar manner.

\subsection{Video Denoising Additional Experiments and Details}
\label{apx:vid}

The link to the video data
from \citep[Section V-A]{li2004statistical}
has been broken for a while,
but the videos can still be found using the internet archive.
We
downloaded the videos from \url{https://web.archive.org/web/20080118111318/http://perception.i2r.a-star.edu.sg/bk_model/bk_index.html}. 

\begin{figure}
    \begin{subfigure}[b]{0.48\textwidth}
         \centering
        \includegraphics[width=\textwidth]{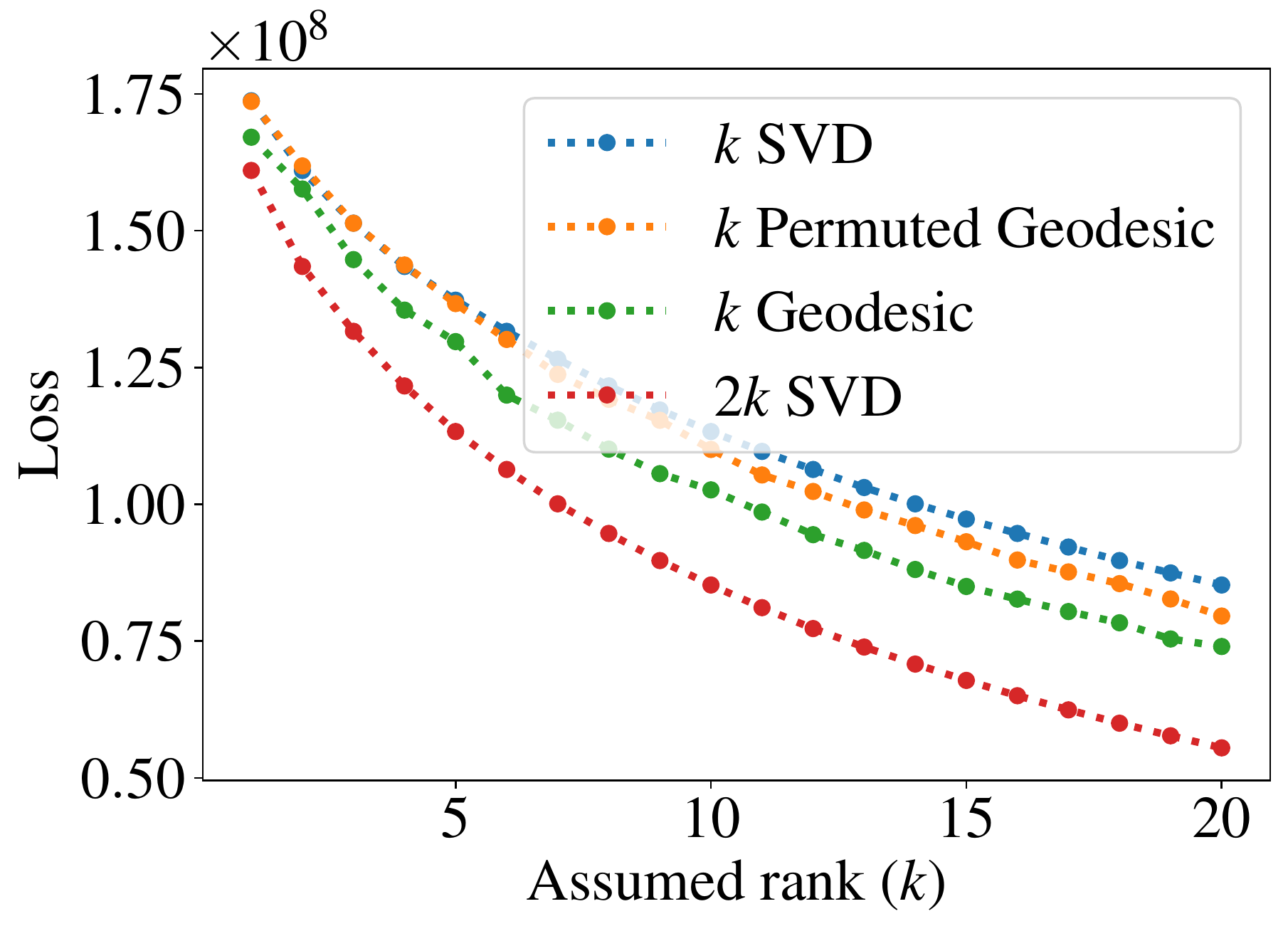}
        \caption{}
        \label{fig:cascade_lossvrank}
     \end{subfigure}
     \hfill
     \begin{subfigure}[b]{0.48\textwidth}
         \centering
        \includegraphics[width=\textwidth]{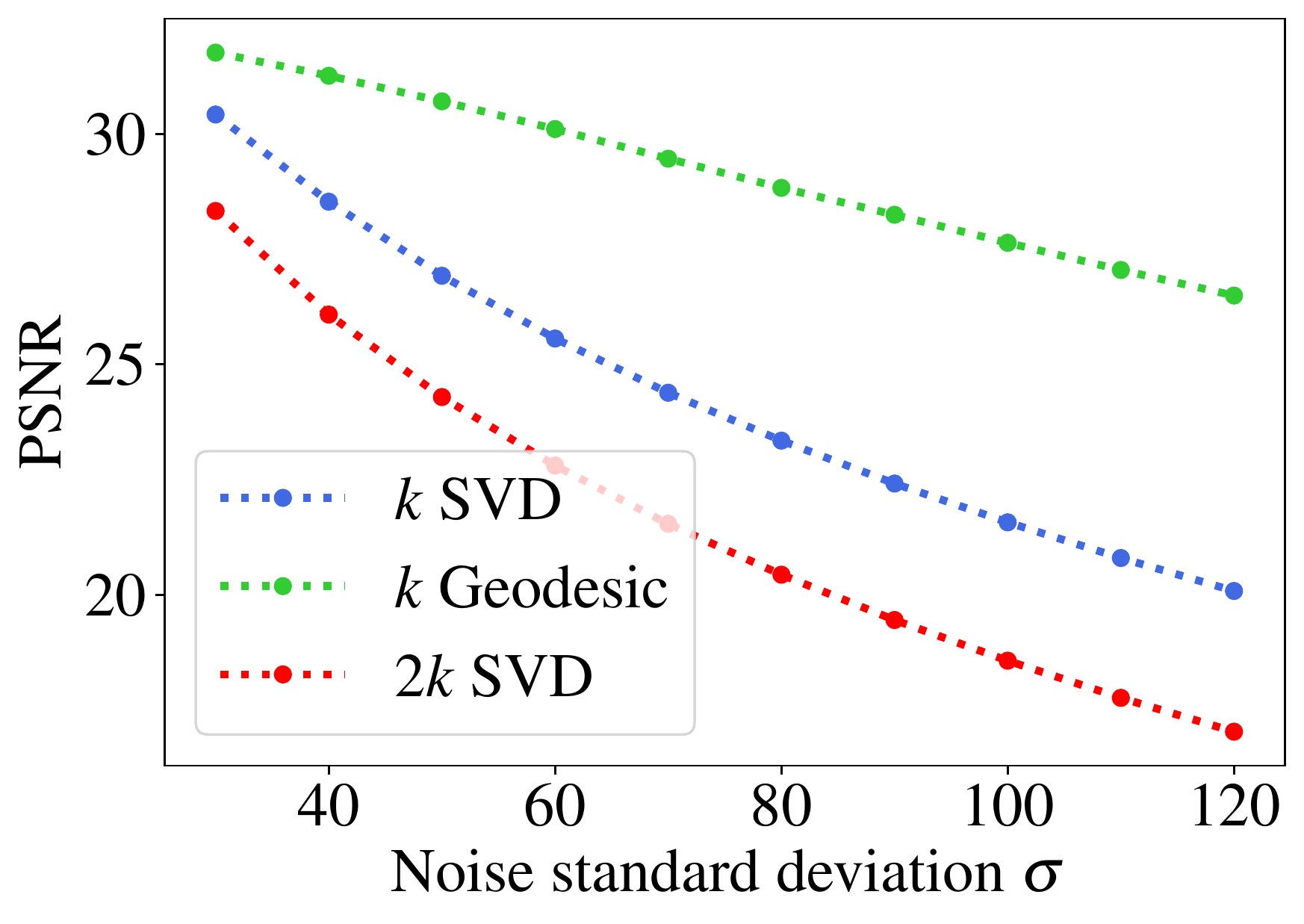}
        \caption{}
        \label{fig:cascade_psnrvnoise}
     \end{subfigure}
      \hfill
     \caption{Quantitative evaluation of geodesic subspace model for waterfall video sequence. In (a) loss from \eqref{eqn:loss2} is plotted for a video sequence containing 260 frames/images. Loss is plotted against different values of assumed rank of data $k$. In (b) we added AWGN to the video data and then applied rank-$k$ SVD, rank-$2k$ SVD, and the geodesic model to denoise the noisy version of video with $k=10$ and $\ell=4$.}
     \label{fig:video_cascade}
\end{figure}

\begin{figure}
\begin{subfigure}{0.48\textwidth}
    \centering
    \includegraphics[width=\textwidth]{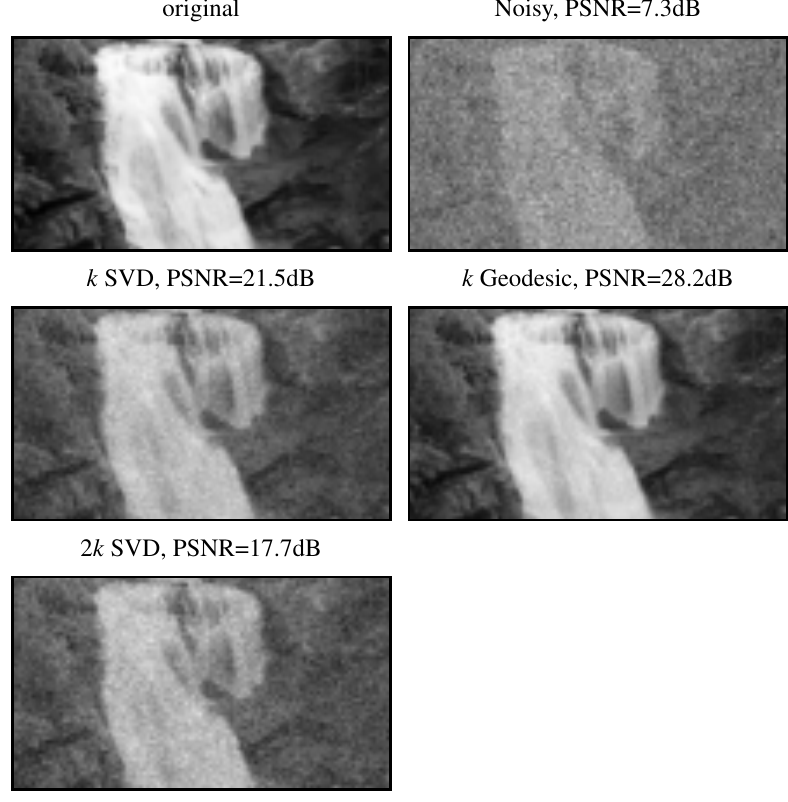}
    \caption{}
    \label{fig:video_waterfall_2}
    \end{subfigure}
\begin{subfigure}{0.48\textwidth}
    \centering
    \includegraphics[width=\textwidth]{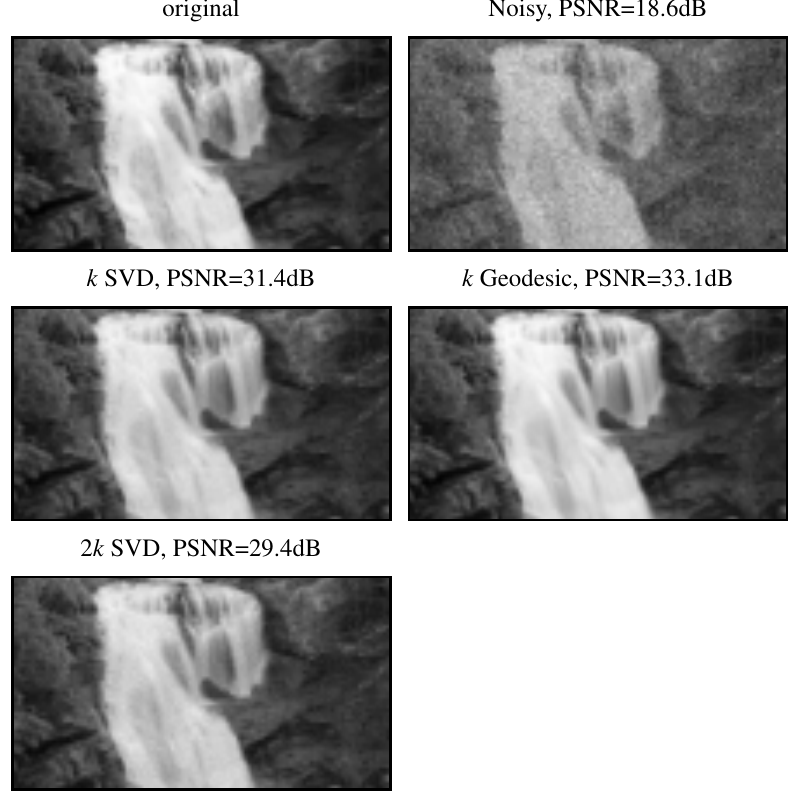}
    \caption{}
    \label{fig:video_waterfall_3}
    \end{subfigure}
    \caption{Visual example of denoising frame 125 in the waterfall video sequence with AWGN of $\sigma=110$ in (a) and $\sigma=30$ in (b). The geodesic model was able to denoise the noisy image more effectively than rank-$k$ SVD and rank-$2k$ SVD in high noise regime in (a). On the other hand for lower noise regime in (b) the denoising performance is quite similar.}
    \label{fig:video_waterfall_visual}
\end{figure}

We provide further denoising results for video data in this section. We perform experiments on waterfall video data (in addition to the curtain video we provided in the main text), and the results are presented in Figure~\ref{fig:video_cascade} and Figure~\ref{fig:video_waterfall_3}. Same as before, we start by showing that the geodesic model is a good choice for this video data. Figure~\ref{fig:cascade_lossvrank}
shows the training loss
as a function of the rank $k$.
The training loss for the geodesic model lies
in between $k$ and $2k$ cases, as expected.
It also has smaller error as compared to permuted data, which is strong evidence that the geodesic model is a reasonable model to consider for the waterfall video sequence.
Next, we study the denoising capabilities of rank-$k$ SVD, rank-$2k$ SVD, and rank-$k$ geodesic model by adding AWGN
with different values of standard deviation $\sigma$ to the video sequence
and applying these three approaches to remove noise. The quality of the denoised image is measured using the peak signal to noise ratio (PSNR).  Figure~\ref{fig:cascade_psnrvnoise} shows the PSNR of the denoised video
as a function of added noise level.
Finally, we provide visual evidence of denoising in Figure~\ref{fig:video_waterfall_visual}. In Figure~\ref{fig:video_waterfall_2} frame 125 is shown for denoising the video corrupted with AWGN of $\sigma=110$. Each image shown is the reconstruction of that frame using each model for denoising, and the PSNR is given at the top of each image. Notice that both PSNR and the perceptual quality of image denoised by the rank-$k$ geodesic model is better than the other two methods, which is a similar trend we observed in Figure~\ref{fig:frame125} for the Curtain dataset. Next, in Figure~\ref{fig:video_waterfall_3} we show results from a similar experiment but the noise added is AWGN with $\sigma=30$, which is significantly lower than the previous experiment. From this experiment we can conclude that the three methods have almost similar performance in low noise settings.

\begin{figure}
\begin{subfigure}{0.42\textwidth}
    \centering
    \includegraphics[width=\textwidth]{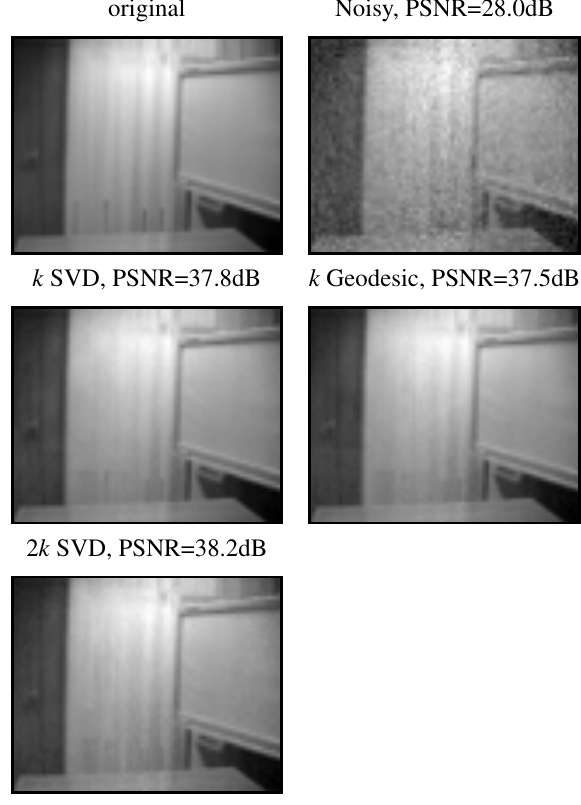}
    \caption{}
    \label{fig:video_curtain_lownoise}
    \end{subfigure}
\begin{subfigure}{0.57\textwidth}
    \centering
    \includegraphics[width=\textwidth]{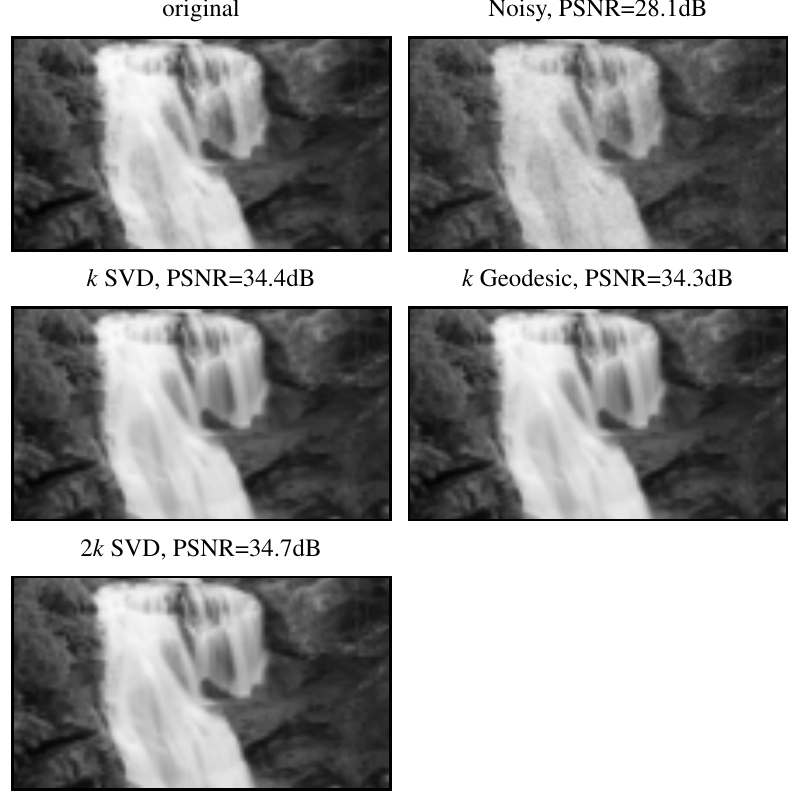}
    \caption{}
    \label{fig:video_waterfall_lownoise}
    \end{subfigure}
    \caption{When the SNR is higher, the geodesic model and PCA model are more comparable.
    This figure shows the same frames from both the curtain and waterfall videos and their reconstructions with $\sigma = 10$.}
    \label{fig:video_high_snr}
\end{figure}

Finally, we display frame 125 for curtain and waterfall sequences
for a lower noise (higher SNR) regime in Figure~\ref{fig:video_high_snr}. These results show that in lower noise settings rank-$2k$ SVD is able to learn more structure in data and hence has better denoising performance than rank-$k$ SVD and rank-$k$ geodesic model. In contrast, in higher noise settings the structure in the images described by smaller singular values is overwhelmed by noise and rank-$2k$ SVD ends up learning a lot of noise, which  diminishes its denoising capabilities.

\end{document}